\documentstyle[emulateapj,epsf,apjfonts]{article}

\newcommand\kms{\ifmmode{~\rm km\th s^{-1}}\else ~km\th s$^{-1}$\fi}
\newcommand\th{\thinspace}
\newcommand\vi{\hbox{$V\!-\!I$}}
\newcommand\vk{\hbox{$V\!-\!K$}}
\newcommand\ri{\hbox{$R\!-\!I$}}
\newcommand\by{\hbox{$b\!-\!y$}}

\lefthead{Torres \& Ribas}
\righthead{YY~Gem}

\begin{document}

\title{Absolute dimensions of the M-type eclipsing binary YY~Geminorum
(Castor~C): a challenge to evolutionary models in the lower
main-sequence\altaffilmark{1}}
 
\altaffiltext{1}{Some of the observations reported here were obtained
with the Multiple Mirror Telescope, a joint facility of the
Smithsonian Institution and the University of Arizona.}

\author{Guillermo Torres}
\affil{Harvard-Smithsonian Center for Astrophysics, 60 Garden St.,
Cambridge, MA 02138}
\authoremail{gtorres@cfa.harvard.edu}

\author{Ignasi Ribas}
\affil{Department of Astronomy \& Astrophysics, Villanova University,
Villanova, PA 19085}
\affil{Departament d'Astronomia i Meteorologia, Universitat de Barcelona,
Av.\ Diagonal, 647, E-08028 Barcelona, Spain}
\authoremail{iribas@ast.villanova.edu}

\begin{abstract}

We present new spectroscopic observations of the detached late-type
double-lined eclipsing binary YY~Gem ($P = 0.814$~days), a member of
the Castor sextuple system and one of the benchmarks for the
comparison between observations and stellar evolution theory in the
lower main-sequence.  In addition, we have re-analyzed existing light
curves in several passbands using modern techniques that account for
the conspicuous presence of spots. This, combined with the
spectroscopy, has yielded a very precise determination of the absolute
dimensions of the components, which are virtually identical to each
other. We obtain for the mean mass, radius, and effective temperature
the values $M = 0.5992 \pm 0.0047$~M$_{\sun}$, $R = 0.6191 \pm
0.0057$~R$_{\sun}$, and $T_{\rm eff} = 3820 \pm 100$~K. Both the mass
and the radius determinations are good to better than 1\%, which in
the case of the radius represents a fourfold improvement over previous
results and significantly enhances the value of this quantity for
testing the models. We discuss the importance of systematic effects in
these measurements, by comparison with another high-precision
determination of the mass by S\'egransan et al.\ (2000).  A
re-analysis of the Hipparcos transit data for Castor~AB that accounts
for the relative motion of the pair in its 467~yr-period orbit has
yielded an improved parallax for the system of $66.90 \pm 0.63$~mas.
With this, we have estimated the age ($\sim$370~Myr) and metal
abundance ([Fe/H]~$\sim 0.0$) of YY~Gem from isochrone fits to Castor
A and B under the assumption of a common origin. This, along with the
other physical properties, allows for an unusually stringent test of
the models for low-mass stars. We have compared the observations of
YY~Gem with a large number of recent theoretical calculations, and we
show that all models underestimate the radius by up to 20\%, and most
overestimate the effective temperature by 150~K or more.  Both of
these trends are confirmed by observations of another similar system
in the Hyades cluster (V818~Tau).  Consequently, theoretical ages for
relatively low-mass objects such as T~Tauri stars derived by placing
them on the H-R diagram may be considerably biased. If the radius is
used directly as a measure of evolution, ages could be underestimated
by as much as a factor of 10 in this mass regime.  In view of these
discrepancies, absolute ages from essentially all current models for
the lower main sequence must be viewed with at least some measure of
skepticism.  Finally, we derive a new and very accurate ephemeris
based on all available times of eclipse, and we lay to rest previous
claims of sudden changes in the orbital period of the binary, which we
show to be spurious. 
	
\end{abstract}


\section{Introduction}
\label{secintro}

The validation of theoretical models of stellar structure and stellar
evolution relies heavily upon the accurate determination of the
absolute dimensions of stars in detached eclipsing binary systems. For
main-sequence stars in the mass range from about 1 to 10~M$_{\sun}$
the models are now reasonably well constrained by the observations,
with some four or five dozen systems available that have the required
precision of 1--2\% in the masses and radii (see
\markcite{a91}Andersen 1991; \markcite{m97}1997). The situation in the
lower main sequence, however, is far less satisfactory.  Despite the
ubiquity of K and M stars in the Galaxy, known cases of detached
eclipsing binary systems in which one or both components are of low
mass are extremely rare because of observational limitations (the
stars are fainter and therefore more difficult to study).  For decades,
only two systems composed of M-type stars were known: CM~Dra
(\ion{M4.5}{5}e, mass $\sim$ 0.25~M$_{\sun}$; \markcite{l77}Lacy 1977;
\markcite{m96}Metcalfe et al.\ 1996) and YY~Gem (\ion{M1.0}{5}e, mass
$\sim$ 0.6~M$_{\sun}$; \markcite{b74}Bopp 1974; \markcite{ls78}Leung
\& Schneider 1978).  Very recently a third example has been found,
CU~Cnc (\ion{M3.5}{5}e; \markcite{d99}Delfosse et al.\ 1999), with
component masses that are intermediate between the other two
systems\footnote{In addition to these, two other eclipsing systems
with M-type components have been discovered by the OGLE microlensing
project (\markcite{u93}Udalski et al.\ 1993) in the direction of the
Galactic bulge: BW3~V38 (\markcite{mr97}Maceroni \& Rucinski 1997) and
BW5~V173 (\markcite{mr99}Maceroni \& Rucinski 1999). The first of
these is a very tight binary that may or may not be detached. The
second appears well detached. However, both are very faint ($V \approx
18.3$ and 20.5, respectively), which makes them exceedingly difficult
to study.}.  All three binaries are flare stars. 

YY~Gem (Castor~C, Gliese~278C, BD$+32\arcdeg1582$, $\alpha = 7^h 34^m
37\fs5$, $\delta = +31\arcdeg 52\arcmin 09\arcsec$, J2000, $V=9.1$)
belongs to a remarkable sextuple system in which the brightest member
is $\alpha$~Geminorum (Castor). $\alpha$~Gem itself is a visual binary
with a current angular separation of about 3\farcs9.  The components
of Castor orbit each other with a period of 467~yr
(\markcite{h88}Heintz 1988). Each of them, in turn, is a single-lined
spectroscopic binary: Castor~A ($\alpha^2$~Gem, HD~60179; the brighter
star) has a period of 9.21~days, and Castor~B ($\alpha^1$~Gem,
HD~60178) has a period of 2.93~days (\markcite{vh40}Vinter-Hansen
1940). YY~Gem is separated from these two by $71\arcsec$, it is at the
same distance and has the same proper motion, and it is a double-lined
spectroscopic binary with a period of 0.814 days (19.5 hours). Its
duplicity was discovered spectroscopically in 1916 by
\markcite{aj20}Adams \& Joy (1920), and the first spectroscopic orbit
was reported by \markcite{js26}Joy \& Sanford (1926). Almost
simultaneously \markcite{vg26}van Gent (1926) discovered eclipses, and
presented a light curve based on photographic observations. Numerous
studies since then have established that the orbit is essentially
circular and that the components are very similar in all their
properties.  Modern photoelectric observations were obtained by
\markcite{ls78}Leung \& Schneider (1978, hereafter LS78), and showed a
complication that earlier observers had also noted, namely,
irregularities or distortions in the light curve attributed to the
presence of surface inhomogeneities (spots) that are common in this
class of stars.  These have been generally dealt with by a variety of
ad-hoc rectification procedures. YY~Gem is in fact the first star,
other than the Sun, on which these maculation effects were validated
(\markcite{k52}Kron 1952).  The spectroscopic study by
\markcite{b74}Bopp (1974), which until now has set the limit on the
precision of the masses at approximately 4\% (see
\markcite{a91}Andersen 1991), has very recently been superseded by a
new study by \markcite{sg00}S\'egransan et al.\ (2000; hereafter S00)
reporting formal errors as small as 0.2\%. 
	
Considerable progress has been made in recent years in the modeling of
low-mass stars (see, e.g., \markcite{s95}Saumon, Chabrier \& Van Horn
1995; \markcite{cb97}Chabrier \& Baraffe 1997;
\markcite{h99}Hauschildt, Allard \& Baron 1999, and references
therein), with significant improvements in the model atmospheres that
are now applied as boundary conditions to the interior equations, as
well as in the opacities, the equation of state, and other
ingredients.  Because of the paucity of suitable binaries in this
regime, the \emph{accuracy} as much as the \emph{precision} of the
fundamental properties derived for the stars in the three known
low-mass eclipsing systems are particularly important if the
comparison with theory is to be helpful. At the level of 1\% in the
uncertainties, as required for such tests, the data must be examined
very carefully for \emph{systematic} errors that might bias the
results (particularly the masses, which hinge on the spectroscopy),
since these often represent the dominant contribution to the
uncertainty. 

The spectroscopic study by \markcite{sg00}S00 certainly meets the
demands for high precision in the masses.  In this paper we present
our own spectroscopic investigation of YY~Gem based on new
observations made at the Harvard-Smithsonian Center for Astrophysics,
yielding masses with formal uncertainties also under 1\%. Thus, two
completely independent spectroscopic studies with similarly small
errors are now available for YY~Gem, a rare occurrence for eclipsing
binaries in general, and unprecedented for the lower main sequence.
This presents us with a unique opportunity to evaluate the consistency
between internal and external mass errors in this key system. 

Accurate determinations of the radii, a sensitive indicator of
evolution, also contribute critical information for the comparison
with theory.  In addition to our new mass determinations, we have
re-analyzed existing light curves from the literature with modern
methods in order to obtain improved estimates of the size of the
stars in YY~Gem that are also good to better than 1\%. 

As we discuss in \S\ref{secdisc}, our results provide particularly
strong constraints on evolutionary tracks and suggest rather
significant shortcomings in recent model calculations that require
careful attention. The impact of these shortcomings is felt especially
in the field of young stars, where the theoretical ages deduced for
T~Tauri stars and other objects occupying the lower regions of the H-R
diagram may be seriously biased. 

The motivation for this paper is thus fourfold: ($i$) To establish
accurate masses for the components of YY~Gem from our new
radial-velocity observations, with a detailed discussion of potential
systematic errors intrinsic to our study and also compared to other
studies; ($ii$) to improve the radii by re-analyzing existing light
curves with modern techniques that can model spots simultaneously with
the other parameters of the light curve, and which were not available
at the time the original photometric studies were made; ($iii$) to
carry out a detailed comparison with recent stellar evolution models
for low-mass stars in order to evaluate competing calculations and to
test the physical assumptions involved; ($iv$) to investigate
unconfirmed claims of changes in the orbital period of YY~Gem that are
quite unexpected for a system of its characteristics, and have
remained unexplained. 
	
\section{Spectroscopic observations and reductions}
\label{secspecobs}

YY~Gem was observed at the Harvard-Smithsonian Center for Astrophysics
(CfA) from 1996 October to 1998 January. We used three nearly
identical echelle spectrographs on the 1.5-m Wyeth reflector at the
Oak Ridge Observatory (Harvard, Massachusetts), the 1.5-m Tillinghast
reflector at the F.\ L.\ Whipple Observatory (Mt.\ Hopkins, Arizona),
and on the Multiple Mirror Telescope (also Mt.\ Hopkins, Arizona)
prior to its conversion to a monolithic 6.5-m mirror. A single echelle
order centered at 5187~\AA~was recorded using intensified
photon-counting Reticon detectors, giving a spectral window of about
45~\AA. The resolving power of these instruments is
$\lambda/\Delta\lambda \approx 35,\!000$, and the signal-to-noise
(S/N) ratios achieved for YY~Gem range from about 10 to 40 per
resolution element of 8.5\kms.  A total of 90 spectra were obtained.
The stability of the zero-point of our velocity system was monitored
by means of exposures of the dusk and dawn sky, and small systematic
run-to-run corrections were applied in the manner described by
\markcite{l92}Latham (1992). 

Radial velocities were derived with the two-dimensional
cross-correlation algorithm TODCOR (\markcite{zm94}Zucker \& Mazeh
1994), which uses a combination of two templates, matched to each of
the components of the binary. This method is ideally suited to our low
S/N observations, and has the important advantage that it minimizes
potential errors caused by ``peak pulling" in the standard
one-dimensional correlations of composite spectra. By this we refer to
systematic shifts in the velocities due to blending between the two
main correlation peaks, or between either peak and the sidelobes of
the other peak. The second of these problems can be significant in
correlations over a narrow wavelength window such as ours, where only
a limited number of spectral lines are available (see
\markcite{l96}Latham et al.\ 1996). 

The selection of templates for the components of YY~Gem presents a
special challenge given the low temperatures of the stars and the
demands on high accuracy for this particular system. In regular binary
work at CfA on systems of earlier spectral type the templates for the
correlations are usually selected from an extensive library of
synthetic spectra computed for us by J.\ Morse, that are based on
model atmospheres by R.\ L.\ Kurucz\footnote{Available at {\tt
http://cfaku5.harvard.edu}.}.  These synthetic spectra provide an
excellent fit to the real observations, and have the advantage that
they can be computed over a wide range of effective temperatures
($T_{\rm eff}$), metallicities ([m/H]), surface gravities ($\log g$),
and projected rotational velocities ($v \sin i$), specifically for the
resolution and wavelength region of our observed spectra (J.\ A.\
Morse \& R.\ L. Kurucz 2002, in preparation; see also
\markcite{n94}Nordstr\"om et al.\ 1994).  However, for temperatures
much cooler than about 4000~K, which is precisely the case for YY~Gem,
the match to the spectra of real stars is degraded mainly because the
models do not incorporate some key molecular opacity sources that
become important in this regime.  Consequently, there is the potential
for systematic errors in the radial velocities and in the derived
masses. As an alternative, templates based on real spectra taken with
the same instrumental setup may be used, but these present
complications of their own, not the least of which is the limited
choice of sufficiently bright single stars with precisely the right
spectral type.  For example, the overlap in wavelength with the
spectra of YY~Gem will generally be reduced somewhat depending on the
Doppler shift of the star selected as the template (in addition to the
difference in heliocentric velocities), whereas the synthetic spectra
described above were calculated over a wider wavelength range
precisely to accommodate relative Doppler shifts.  We have performed
extensive series of tests with both kinds of templates, as described
in more detail below. 

Given the similarity of the components in YY~Gem, initially we
selected synthetic templates with an effective temperature of 3750~K
for both stars, based on preliminary estimates derived from
photometry.  The surface gravity was set to $\log g = 4.5$, and we
adopted the solar metallicity. To determine the appropriate rotational
broadening we ran grids of cross-correlations with TODCOR for a wide
range of values of $v \sin i$ separately for each template. We
selected the combination giving the highest correlation, averaged over
all exposures and weighted by the corresponding S/N ratio of each
observation. 

For historical reasons, and to avoid confusion in the identification
of the nearly identical components, we follow in this paper the naming
convention adopted by \markcite{vg31}van Gent (1931) based on the
appearance of the light curve. The star eclipsed at the slightly
deeper minimum (phase 0.0, Min~I) is referred to as the ``primary", or
star ``A", in the standard photometric usage. The epoch of this
minimum is given in \S\ref{secperiod}, and differs only by an integer
number of cycles from the original epoch reported by van Gent. As it
turns out, the mass of star~A is formally smaller than that of the
other component, though only marginally so. 

\begin{figure}
\figurenum{1}
\epsscale{0.50}
\plotone{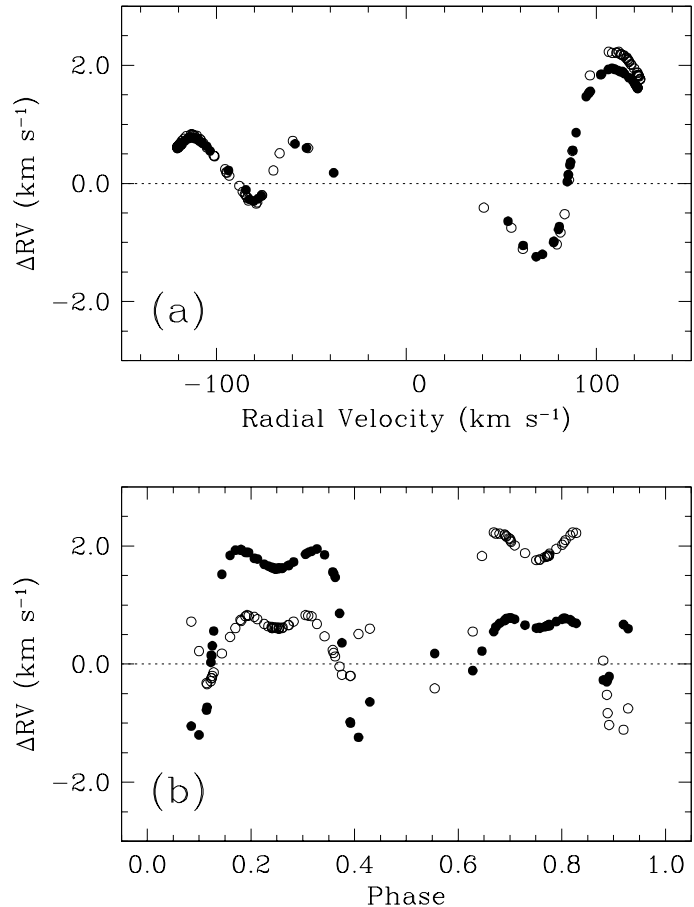}
\caption{Systematic errors in the measured radial
velocities of YY~Gem due to line blending, from simulations with
synthetic binary templates. The photometric primary is represented by
open circles. The differences, shown as a function of radial velocity
(a) and phase (b), were applied to the measured velocities as
corrections. \label{figtodcorcor}}
\end{figure}

Residual systematic errors in the velocities resulting from the narrow
spectral window were checked by means of numerical simulations as
discussed in detail by \markcite{l96}Latham et al.\ (1996). Briefly,
we generated a set of artificial binary spectra by combining the
primary and secondary templates in the appropriate ratio and applying
velocity shifts for both components as computed from a preliminary
orbital solution at the actual times of observation of each of our
spectra. These artificial spectra were then processed with TODCOR in
exactly the same way as the real observations, and the resulting
velocities were compared with the input (synthetic) values. The
differences (TODCOR minus synthetic) are shown in
Figure~\ref{figtodcorcor} as a function of velocity and also orbital
phase, and are clearly seen to be systematic in nature. They reach
values as large as $2\kms$ at the phase of maximum velocity
separation, or about 1.5\% of the velocity semi-amplitudes
($\sim$120\kms). This translates into a 5\% effect in the masses,
which is very significant. These differences were applied as
corrections to the velocities measured from the real spectra of
YY~Gem. 

Experiments were carried out using the spectrum of a real star as the
template, instead of calculated spectra.  For this we chose Gliese~908
(BD$+01\arcdeg$4774, HIP~117473), which has a spectral type of
\ion{M1}{5}e (\markcite{h94}Henry, Kirkpatrick \& Simons 1994),
matching that of YY~Gem. A strong exposure of Gliese~908 was obtained
for this purpose with our instrumentation at the Multiple Mirror
Telescope. The spectrum of the star was rotationally broadened by
various amounts, and grids of correlations were run in the same way as
above to determine the values that best match each component of
YY~Gem. Radial velocities were obtained with TODCOR as before, and
corrections were determined by creating artificial composite spectra
based on Gliese~908. The results were generally not as satisfactory as
with the synthetic spectra.  Correlation values were typically lower,
the residuals from the orbital solution were somewhat larger and
showed systematic trends, and the corrections for systematic effects
based on simulations were some 3 times larger. The latter suggested to
us that the use of this observed template carries a higher risk of
residual systematic errors.  In addition, the light ratio came out
significantly smaller than unity, a rather unexpected result for stars
of nearly equal temperature and mass (see below). 

Based on these experiments we chose to adopt synthetic templates for
the cross-correlations, with final parameters adjusted slightly to
$T_{\rm eff} = 3820$~K and $v \sin i = 37\kms$ for both stars.  The
temperature of the mean component is based on a number of photometric
indices (see \S\ref{secabsdim}), while the projected rotational
velocity is the average of the individual determinations (formally
36.4\kms~and 37.8\kms~for star~A and B, respectively, with estimated
uncertainties of 2.0\kms).  Templates with these parameters were
calculated by interpolation in our library of synthetic spectra.  We
repeated the procedure described above to derive the corrections for
systematic errors due to the narrow spectral range, which changed very
little. 

The light ratio was obtained from our spectra following the procedure
outlined by \markcite{zm94}Zucker \& Mazeh (1994). Numerical
simulations analogous to those described above were carried out to
correct for systematic effects that result as the spectral lines of
the components shift in and out of the spectral window as a function
of phase. These errors reach a maximum of about 15\% at the phases of
maximum velocity separation between the components. The value we
obtained, after corrections, is $L_B/L_A = 1.088 \pm 0.047$. Strictly
speaking, this is the ratio at the mean wavelength of our spectra
(5187~\AA), but, because the stars are so similar, for all practical
purposes it can be considered to be the same in the visual band. 

The final heliocentric radial velocities we obtained, including
corrections, are listed in Table~\ref{tabrvs}. As will be seen later
in \S\ref{seclightcurves}, the stars in YY~Gem are so nearly spherical
that the effects of tidal and rotational distortions on the measured
radial velocities are negligible for this system. 

\begin{figure}
\figurenum{2}
\epsscale{0.80}
\plotone{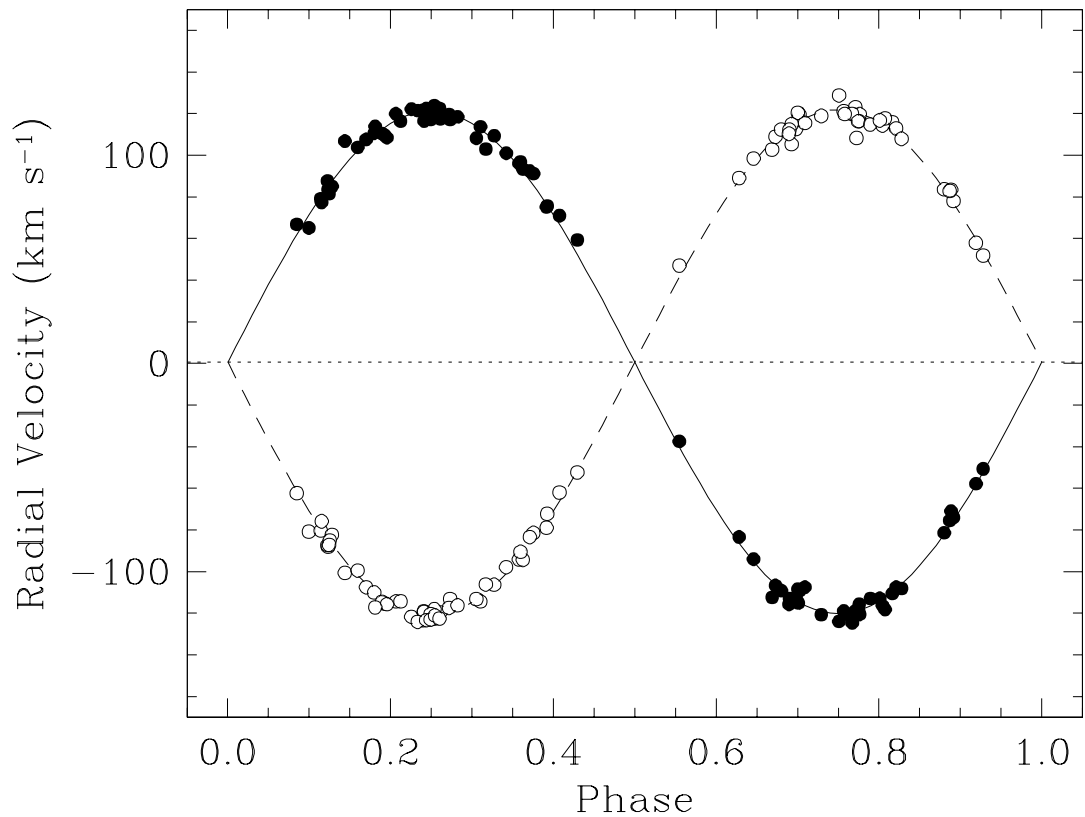}
\caption{Radial velocity measurements for YY~Gem
and corresponding velocity curves. Open circles represent the
photometric primary.\label{figspecorb}}
\end{figure}

Our spectroscopic orbital solution is shown graphically in
Figure~\ref{figspecorb}, along with the observations. The ephemeris
adopted here is that of \S\ref{secperiod}, based on an analysis of all
available times of minimum. The elements of the orbit are given in
Table~\ref{tabspecelem}, and the residuals for each velocity are
listed with the observations in Table~\ref{tabrvs}. Tests allowing for
a non-circular orbit indicated an eccentricity that was not
significantly different from zero ($e = 0.0034 \pm 0.0032$), in
agreement with evidence described later from the times of minimum
(\S\ref{secperiod}) and from the light curve (\S\ref{seclightcurves}).
Consequently, we adopted a circular orbit. 
	
In a very recent paper \markcite{sg00}S00 reported new spectroscopic
observations of YY~Gem obtained with the ELODIE spectrograph
(\markcite{b96}Baranne et al.\ 1996), a high-precision instrument used
for extrasolar planet searches. Due to the relatively rapid rotation
of the components, the formal uncertainty of a single observation of
YY~Gem in this new study is degraded with respect to the normal
performance of that instrument for sharp-lined stars, so that the
precision of an individual radial velocity measurement turns out to be
roughly the same as ours. However, the orbital fitting procedure
adopted by \markcite{sg00}S00 is different from ours in that they
incorporate the one-dimensional cross-correlation profiles directly,
rather than first deriving radial velocities and then solving for the
orbital elements.  The velocity semi-amplitudes that they
obtain\footnote{The values reported originally by \markcite{sg00}S00
are incorrect, as we describe in more detail in \S\ref{secmasses}. The
semi-amplitudes we give here are the result of a more sophisticated
adjustment and are to be preferred (Forveille 2001, priv.\ comm.)}
($K_A = 121.87 \pm 0.12\kms$ and $K_B = 121.24 \pm 0.11\kms$), which
ultimately set the masses, are nevertheless quite similar to ours (see
Table~\ref{tabspecelem}).  The main difference is that those values
are systematically larger than ours by about $0.7\kms$, a discrepancy
of $\sim1.5\sigma$. We discuss the significance of this difference for
the mass determinations in \S\ref{secabsdim}.  Their mass ratio ($q
\equiv M_B/M_A = 1.0052$) is very similar to the value we
derive. 

\section{The orbital period of YY~Gem}
\label{secperiod}

Numerous photometric and spectroscopic studies of the system over the
years have been based on a period identical to or close to that
derived by \markcite{vg31}van Gent (1931) in his original photographic
investigation of YY~Gem, which is $P = 0.8142822 \pm 0.0000010$~days.
No evidence of any change in this period was reported in the
literature until the study by \markcite{ls78}LS78, who observed the
star photoelectrically in 1971 and carried out a detailed light-curve
analysis. These authors claimed that the classical period did not fit
their data, and derived a much longer period of $P = 0.81679 \pm
0.00046$~days from four times of minimum based on their own
observations (one primary minimum and three secondary minima).  This
very significant change in the period ($\sim$5.5$\sigma$) is difficult
to understand for normal main-sequence stars that are not interacting,
and prompted \markcite{m80a}Mallama (1980a) to investigate the issue.
He concluded that the \markcite{ls78}LS78 period was most likely in
error, probably because of the very short baseline of those
observations (only six days, with the eclipse timings spanning only
four days).  \markcite{ki80}Kodaira \& Ichimura (1980) also studied
the problem, and again dismissed the period of \markcite{ls78}LS78 as
erroneous, ascribing it to irregularities in the 1971 light curve due
to spots, which could have affected the times of minimum derived by
Leung \& Schneider.  Neither of these explanations is very easy to
accept given the relatively high quality and completeness of the
observations by \markcite{ls78}LS78, and this discrepancy in the
period has remained something of a mystery ever since. 

We have re-examined the issue here, in an effort to provide an
accurate understanding and a new determination of the period of YY~Gem
for our own re-analysis of the \markcite{ls78}LS78 and other light
curves in \S\ref{seclightcurves}.  In Table~\ref{tabtmin} we have
collected from the literature all available times of minimum in order
to look for any trends suggesting changes in the period. Excluding the
four measurements by \markcite{ls78}LS78, there are 57 determinations
for the primary and 55 for the secondary spanning more than 73 years,
obtained using a variety of methods.  One of them (a secondary minimum
indicated in parenthesis in Table~\ref{tabtmin}) gave an unusually
large residual, and was rejected.  We performed a least-squares fit
for a linear ephemeris using these data, with errors adopted from the
original sources when available. Most determinations have no published
errors, however, and we determined these iteratively based on the mean
residuals for each type of observation (photographic, visual, and
photoelectric) from preliminary fits.  The adopted uncertainties for
each method are $\sigma_{\rm pg} = 0.0021$ days, $\sigma_{\rm v} =
0.0080$ days, and $\sigma_{\rm pe} = 0.0018$ days, respectively. No
unusual patterns were seen in the residuals.  The ephemeris derived is
 \begin{equation}
{\rm Min~I~(HJD)} = 2,\!449,\!345.112327(87) + 0.814282212(12)\cdot E~,
\end{equation}
 in which the period is nearly identical to that by \markcite{vg31}van
Gent (1931), but two orders of magnitude more precise. Thus, there is
no evidence of any significant change in the period from the eclipse
timings prior to, or following those of \markcite{ls78}LS78. 

The photoelectric observations by \markcite{ls78}LS78 were obtained in
1971 January 13-19 at the Kitt Peak National Observatory. The authors
list the heliocentric Julian date and magnitude of YY~Gem in the
$V\!RI$ filters for each of their measurements. Using the new
ephemeris above we have confirmed the claim by Leung \& Schneider that
their light curve is \emph{not} well fit by the classical period, and
that the longer value they report is clearly better for their data.
Barring a sudden change in period with a subsequent return to the
previous value ---a highly unlikely scenario--- this indicates that
the problem must lie in the times of observation recorded by
\markcite{ls78}LS78. 

The average residual of their four times of minimum from the new
ephemeris is $+$0.0038 days. Though somewhat larger than typical for
photoelectric determinations, a more serious concern is a systematic
trend clearly visible over the four consecutive nights such that the
first residual is $-$0.0012 days and the last is $+$0.0079 days,
unacceptably large for the photoelectric technique. In an attempt to
understand this we reconstructed the circumstances of each photometric
observation (time during the night, hour angle, air mass, etc.) based
on the Julian dates tabulated by \markcite{ls78}LS78. In doing so we
noticed that, for the date of these measurements, the Universal Time
(from which the fractional Julian dates are computed) is numerically
very similar to the Local Mean Sidereal Time. A confusion between
these two would explain why the period by \markcite{ls78}LS78 is
longer than the true value, since its units would then be sidereal
days instead of mean solar days. Correcting the \markcite{ls78}LS78
period by the appropriate factor brings it into agreement with our
value in eq.(1) within about 1.5$\sigma$.  Thus, it appears quite
plausible that all Julian dates by \markcite{ls78}LS78 are indeed
affected by this error. 

Once the four eclipse timings of \markcite{ls78}LS78 are corrected,
the increasing trend we noted above in the residuals from the new
ephemeris disappears completely.  However, the average residual is
even larger than before ($-$0.0126 days), pointing perhaps to some
other problem.  The fact that this difference is approximately twice
the value of the heliocentric correction for these dates (which
happens to be near its maximum) may not be entirely a coincidence, and
it is conceivable that \markcite{ls78}LS78 mistakenly applied the
correction with the wrong sign. Under this assumption we tentatively
corrected all the original times of observation, and the average
residual of the four times of minimum was reduced to $-$0.0016, or
about 2 minutes.  This is now essentially the same as the mean error
$\sigma_{\rm pe}$ found above for photoelectric eclipse timings from
other sources. 

With the original \markcite{ls78}LS78 times of observation corrected
for these two effects, the scatter at the minima in their $V\!RI$
light curves when folded with the period in eq.(1) is now marginally
\emph{smaller} than when using their own period determination. That
this is a fortunate coincidence seems very unlikely, and while the
confusion with the sign of the heliocentric correction is perhaps
debatable, it seems clear that the long-standing mystery of the
discrepancy in the period of YY~Gem is adequately explained by the
inadvertent use by Leung \& Schneider of local sidereal times instead
of universal (mean solar) times. 

Because of lingering uncertainties about the second of the corrections
applied to these data, we have chosen not to use the times of minimum
by \markcite{ls78}LS78 as modified above in our final period
determination. Their inclusion has a minimal effect in any case.
Nevertheless, we list them in Table~\ref{tabtmin} for completeness,
with the residuals in parentheses. 

The ephemeris in eq.(1) was derived assuming a circular orbit for
YY~Gem. An independent fit of the times of eclipse allowing for an
eccentric orbit indicated that the secondary minimum occurs at phase
$\Phi_{\rm II} = 0.50018 \pm 0.00014$, essentially consistent with
zero eccentricity.  This agrees with the evidence from spectroscopy
(\S\ref{secspecobs}), and similar indications from the light curve
solutions described below.  Additional fits for separate orbital
periods based on the primary and secondary minima indicated no
detectable apsidal motion: $P_{\rm Min~I} = 0.814282232(18)$~days,
$P_{\rm Min~II} = 0.814282194(19)$~days, with a difference at the
1.5$\sigma$ level. 

\section{Photometric observations and analysis}
\label{seclightcurves}

Light curves for YY~Gem have been obtained by a number of authors
using a variety of techniques since its discovery as an eclipsing
binary. In nearly all cases they have been found to be affected to
some extent by distortions and irregularities that are most likely due
to spots. Among the most complete and precise series of observations
published are those by \markcite{k52}Kron (1952) and
\markcite{ls78}LS78. In both cases the original investigators
rectified the light curves prior to the analysis by removing the
effect of the spots empirically, since the computational methods
available at the time did not allow the irregularities to be taken
into account as an integral part of the analysis. Modern computer
codes have improved considerably over the years incorporating better
physical models, model atmospheres, and also the ability to solve for
the parameters describing the effects of spots on the light curves
simultaneously with the rest of the light elements.  These
improvements can potentially have some effect on the determination of
the radius and other parameters, and have motivated us to re-analyze
the original data sets in an effort to extract the most accurate
results possible. 

The data by \markcite{k52}Kron (1952) consist of 251 photoelectric
observations made with a 0.3-m refractor at the Lick Observatory from
February to April of 1948, at an effective wavelength of 5200~\AA.
Distortions in the light curve due to spots are quite obvious and
appear to affect all phases. Observations were also obtained in the
near infrared ($\lambda\sim 8100$~\AA), but unfortunately those
measurements were never published. The data by \markcite{ls78}LS78
comprise some 790 photoelectric observations in each of the Johnson
$V\!RI$ filters obtained on a 0.9-m telescope at the Kitt Peak
National Observatory during six nights in January of 1971. The
precision (mean error) of an individual measurement is quoted as being
0.012~mag, 0.009~mag, and 0.009 mag in the visual, red, and infrared
bands, respectively.  Once again the light curves show obvious
distortions, particularly in the vicinity of the secondary minimum,
which is slightly asymmetric.  Close inspection also showed certain
phase intervals with significantly larger scatter. Further
investigation revealed that the photometric observations on some
nights were pursued to very high airmass (approaching 3 in the most
extreme case). After careful evaluation of the agreement among the
different nights, we chose to reject observations acquired at
airmasses larger than 1.8, at which we begin to see obvious
distortions.  In addition, a few isolated measurements that appeared
markedly discrepant were eliminated. The number of photometric
observations selected for further analysis was reduced to 640, 635,
and 634 for $V\!RI$, respectively. The ephemeris used to phase all
light curves is that given by eq.(1), and in the case of
\markcite{ls78}LS78 the published times of observation were corrected
before use as described in \S\ref{secperiod}. 

The fits to the light curves were performed using an improved version
of the Wilson-Devinney program (\markcite{WD}Wilson \& Devinney 1971;
hereafter WD) that includes a model atmosphere routine developed by
\markcite{m92}Milone, Stagg \& Kurucz (1992) for the computation of
the stellar radiative parameters, based on the ATLAS9 code by R.\ L.\
Kurucz. A detached configuration, as suggested by the shape of the
light curve (i.e., relatively narrow and similar eclipses) was chosen
for all solutions. Both reflection and proximity effects were taken
into account, even though the light curves do not show strong evidence
for these.  The bolometric albedo was initially fixed at the canonical
value of 0.5 for stars with convective envelopes. The
gravity-brightening coefficient was set to a value of 0.2, following
the theoretical results of \markcite{c00}Claret (2000) for stars with
the effective temperature of YY~Gem. For the limb darkening we used a
logarithmic law as defined in \markcite{ks70}Klinglesmith \& Sobieski
(1970), with first- and second-order coefficients interpolated at each
iteration for the exact $T_{\rm eff}$ and $\log g$ of each component
from a set of tables computed in advance using a grid of Kurucz model
atmospheres.  Negligible differences were seen in the results when
using linear, quadratic, or square-root limb-darkening approximations
for the temperature and surface gravity regime of YY~Gem. 

A mass ratio of $q=M_B/M_A=1.0056$ was adopted from the spectroscopic
solution, and the temperature of the primary component (eclipsed at
phase 0.0) was set to 3820~K, as discussed in \S\ref{secabsdim}. The
iterations in the WD code were carried out in an automatic fashion
until convergence, and a solution was defined as the set of parameters
for which the differential corrections suggested by the program were
smaller than the internal probable errors on three consecutive
iterations.  As a general rule, several sets with different starting
parameters were used to test the uniqueness of the solutions and to
make more realistic estimates of the uncertainties. 

The solutions to the light curves of Kron and \markcite{ls78}LS78 were
run independently because of the conspicuous changes in the location
and geometry of the surface inhomogeneities. We discuss first the
analysis of the light curves of \markcite{ls78}LS78, which span a
wider wavelength range and display better phase coverage. 

\subsection{Fits to the light curves of Leung \& Schneider (1978)}
\label{secls78}

Simultaneous WD fits to the $V\!RI$ light curves were carried out
using the mean error of a single measurement quoted by
\markcite{ls78}LS78 as the relative weight, $w_{\lambda}$, of each
passband (WD ``curve-dependent'' weighting scheme).  For the weighting
of the individual measurements within each passband we experimented
with three different schemes: equal weights for all observations,
double weight to observations within the eclipse (to emphasize the
phases that contain the most information), and photon-noise dependent
weights. Our tests yielded nearly identical solutions (to within
0.2\%) regardless of the weighting scheme used, and in the end we
adopted equal weights for simplicity and objectiveness. 

\begin{figure}
\figurenum{3}
\epsscale{0.70}
\plotone{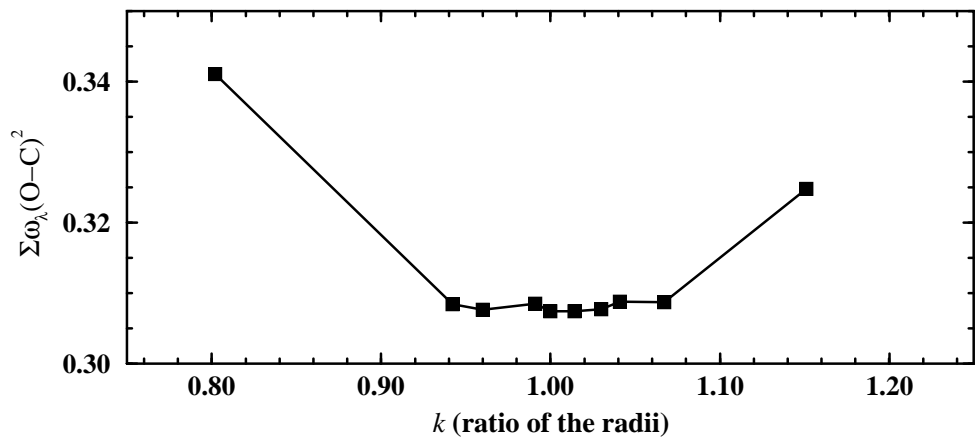}
\caption{$O\!-\!C$ diagram showing the
insensitivity of the Leung \& Schneider (1978) light curve solutions
to the ratio of the radii, $k\equiv r_B/r_A$, for values close to
unity.\label{figk}}
\end{figure}

In our initial WD solutions we solved for the following light curve
parameters: the orbital inclination ($i$), the temperature of the
secondary (or in practice the ratio $T_{\rm eff}^B/T_{\rm eff}^A$),
the gravitational potentials ($\Omega_A$ and $\Omega_B$), the
luminosity of the primary ($L_A$), a phase offset ($\Delta\phi$), and
the spot parameters.  Consistent with earlier indications, the
eccentricity was set to zero to begin with.  Numerous test solutions
revealed early on that the ratio of the radii of the components
($k\equiv r_B/r_A$, where $r_A$ and $r_B$ are the fractional radii in
units of the separation) was poorly constrained. This is not at all
unexpected in systems like YY~Gem, with very similar components that
undergo partial eclipses. In such cases the solutions are often
degenerate because the light curve residuals are insensitive to
variations in $k$.  To evaluate this effect using WD we ran a number
of solutions in which we fixed $\Omega_A$ to values between 6.75 and
7.80 and solved for $\Omega_B$ (probing values of $k$ between $0.80$
and $1.15$). A plot of the residuals as a function of $k$ is shown in
Figure~\ref{figk}. As can be seen, the total weighted sum of residuals
squared, $\sum w_{\lambda}(O\!-\!C)^2$, hardly changes for $k$ in the
range between $0.95$ and $1.05$, leaving the solution essentially
dominated by numerical noise. The rest of the orbital and physical
properties of the system ($i$, $T_{\rm eff}^B/T_{\rm eff}^A$, the sum
of the fractional radii, etc.) remained nearly constant within this
$k$ interval.  Thus, the photometry for YY~Gem is unable to
discriminate between the sizes of the components. 

An alternative is to use the light ratio from spectroscopy as an
external constraint.  However, as we discuss below, the light ratio in
this system is not well defined either, due to the presence of spots.
Under these circumstances, and given that the two stars have nearly
identical masses (to within 0.5\%), we chose to adopt $k=1$ (i.e.,
identical radii) for all light curve solutions.  In practice, the ratio
$k$ is not one of the parameters considered explicitly in the WD code.
Instead, it is the surface potentials of the stars (which, along with
$q$, control the sizes) that can be fixed or left free. To circumvent
this problem we ran solutions by fixing $\Omega_A$ to different values
and leaving $\Omega_B$ free until the resulting fractional radii were
the same.  We found that this occurred for $\Omega_A=7.315$, which
leads to a solution with $\Omega_B=7.334$. For the remainder of the
solutions we fixed $\Omega_A$ to this value of $7.315$. 

It is quite obvious from the sinusoidal shape of the light curves
outside of eclipse that at least one of the components, or perhaps
both, have surface inhomogeneities. This is consistent with the level
of activity displayed by the system (see, e.g., \markcite{mb71}Moffett
\& Bopp 1971; \markcite{fb76}Ferland \& Bopp 1976), but it complicates
the analysis significantly.  We describe the details of our spot
analysis below. 

\subsubsection{Spot modeling}
\label{secspots}

The WD program incorporates a simple spot model with circular spots
that are described by four parameters, namely, the longitude,
latitude, angular radius, and temperature ratio (spot relative to
photosphere). As is well known, the effects of some of these
parameters on the light curves are typically rather similar so that
they can rarely be separated on the basis of the photometric
observations alone, unless the quality of the curves is superb and the
wavelength coverage is very broad. For example, the latitude of a spot
is very weakly constrained by the light curves, and the radius and the
temperature ratio are very strongly correlated.  Furthermore, the
addition of these new parameters generally makes convergence more
problematic, and one is often forced to consider multiple parameter
subsets in order to render the problem manageable (see
\markcite{wb76}Wilson \& Biermann 1976). 

To facilitate the analysis, we have used external constraints to fix
some of the spot parameters.  Doppler imaging is a powerful technique
that employs fits to stellar lines to map asymmetries on the stellar
surface. This technique was successfully applied to YY~Gem by
\markcite{h95}Hatzes (1995) using echelle spectroscopy obtained during
late 1992 and early 1993. The results of Hatzes' analysis indicate
that the spot activity is located within a band centered at a latitude
of about $+45\arcdeg$ on both components (although symmetry about the
equator arises for systems with the high inclination of YY~Gem).  In
addition, the temperature difference between the coolest regions and
the photosphere was found to be between 300~K and 500~K, with the
spots being cooler (dark). Since YY~Gem is a fairly active star we
cannot assume that the spot parameters (including spot size and
longitude) remain constant over the 22-year span between the epoch of
the \markcite{ls78}LS78 photometry and the time of Hatzes' (1995)
spectroscopy.  It seems justified, however, to assume the temperature
contrast of the active regions and their marked confinement in
latitude to be more intimately related to the nature of YY~Gem. There
is, in fact, theoretical evidence that this may be the case.
\markcite{gs00}Granzer et al.\ (2000) have modeled the dynamics of
magnetic flux tubes in stars with masses between 0.4~M$_{\sun}$ and
1.7~M$_{\sun}$, and studied the emergence patterns of star spots at
the stellar surface. For stars with the mass, rotation rate, and
evolutionary state of YY~Gem they predict spot emergence at
intermediate latitudes, just as observed by \markcite{h95}Hatzes
(1995).  Thus, for the light curve analyses in this paper we have
placed the spots at a fixed latitude of $+45\arcdeg$ and assumed them
to be about 400~K cooler than the photosphere (i.e., $T_{\rm
spot}/T_{\rm phot}=0.90$)\footnote{In principle bright spots (plages)
could also explain the observed distortions in the light curves, and
in fact \markcite{k52}Kron's (1952) original analysis called for
bright patches in addition to darker regions. However, it is generally
not possible to discriminate unambiguously between the two kinds of
spots solely on the basis of a light curve. Even the Doppler imaging
technique has similar difficulties unless the spectroscopic
observations are supported by simultaneous multi-wavelength light
curves (see, e.g., \markcite{rs00}Rice \& Strassmeier 2000). In any
case, the effect on the geometric parameters of the system is very
small.  For this reason, and because the features considered in most
previous analyses and also modeled tomographically by
\markcite{h95}Hatzes (1995) are dark, we assume here that the spots
are cooler than the photosphere.}.  The only spot parameters left to
determine are therefore the size and longitude. 

We initially attempted to fit the light curves by considering the
simplest model of a single spot on one of the components. The
solutions converged towards a spot of $26\fdg8$ in radius which
transits the meridian at phase 0.656. However, the light curve fits
displayed obvious systematic patterns in the residuals along the
phases where the spot is in view. Further analysis indicated that the
systematics were a consequence of the inability of the adopted spot
model to fit a wave that extends for almost three quarters of a
period. Intuitively one would expect a small-sized spot to be able to
account for a wave spanning roughly half a period. When both the
latitude of the spot and the temperature contrast were allowed to
vary, the solution tended towards a larger spot (radius of
approximately $55\arcdeg$) at lower latitude ($\approx 35\arcdeg$) and
with a temperature only 90~K cooler than the photosphere. This
(somewhat unphysical) combination of parameters was the code's best
attempt to account for the phase extent of the wave with a single
spot, but resulted in a fit that was rather unsatisfactory.  Tests
with the spot located on either component were run, but no noticeable
differences were seen; the information contained in the light curves
is insufficient to tell which component the spot is located on. 

Given the residual trends seen with a single-spot model, a better and
perhaps more realistic approach to reproducing the observed light
curves would be to consider an elongated spot distribution, especially
since the features actually observed by \markcite{h95}Hatzes (1995)
are in fact extended in longitude.  This was done by adopting two
spots in the WD model located at the same latitude, but at different
longitudes and allowing also for different sizes.  Tests were first
run by assuming both spots to be located on the photometric secondary
(more massive) component. The fit yielded spot radii of $21\fdg0$ and
$23\fdg6$, with meridian transits at phases 0.733 and 0.594 (i.e.,
$50\arcdeg$ apart in longitude), respectively.  This spot
configuration provides a rather good fit to the out-of-eclipse wave in
all filters. Further tests with one spot on each star gave a
marginally better fit (total sum of residuals squared of 0.3095 vs.\
0.3118). The best-fitting spots have radii of $20\fdg1$ and $21\fdg0$,
with meridian transits at phases 0.733 and 0.587, located on the
primary and secondary components, respectively. 

Interestingly, a spectroscopic study of YY~Gem by \markcite{b74}Bopp
(1974) recorded a number of flares from both components, one of which
occurred just three weeks after the observations by
\markcite{ls78}LS78.  Various geometrical restrictions allowed Bopp to
constrain the location of the flaring region to be within $45\arcdeg$
in longitude of phase 0.63. This active region was found to be on the
surface of the secondary star.  Although a direct connection between
this flaring event and the dark spot we find on the secondary
component from our own light curve analysis is difficult to
demonstrate, we note that their phase locations (0.63 and 0.59) are
strikingly similar and are at least consistent with association. 

The need for two spots to adequately reproduce the shape of the
out-of-eclipse region around phase 0.65 seems justified from the
discussion above. Subsequent inspection of the residuals from the fit,
however, revealed the presence of residual systematic deviations near
the first quadrature (0.25) that suggested the presence of yet another
spot. Solutions with a second dark spot on the primary component
indicated that the fit did in fact improve somewhat (total weighted
sum of residuals squared of $0.3076$), and gave a smaller spot with a
radius of $9\fdg0$ that transits the meridian at phase 0.25.  Although
the statistical significance of this additional spot on the primary is
weaker, once again there is independent evidence from flaring activity
that may support its reality. \markcite{b74}Bopp (1974) reported a
series of flare events during late March 1971 and concluded that they
originated from the surface of the primary star in a region centered
around phase 0.22, very close to the location of the dark spot
suggested by the light curves. Based on this we have retained the
spots described above in all subsequent fits of the
\markcite{ls78}LS78 light curves. A graphical representation of the
location of these surface features is presented later in
\S\ref{seckron}, for comparison with the configuration seen 23 years
earlier in the \markcite{k52}Kron (1952) observations. 

As a final comment on the spot fitting procedure it should be pointed
out that the fitting of the out-of-eclipse wave is, in a sense, an
aesthetic exercise that does not really affect the physical and
orbital properties of the YY~Gem system in any significant way
(changes are typically within 1$\sigma$). This partly explains the
strong degeneracy between some of the spot parameters and the
ambiguity as to which component the spot(s) are located on. The proper
fitting of the branches of the minima (particularly at the secondary
eclipse) is somewhat more important.  But spot adjustment is not
without the risk of over-interpreting the data.  The spot distribution
that we have finally adopted is supported to some extent by evidence
from flare studies and provides excellent fits to the light curves.
The solution is not unique, however, and other configurations may be
equally valid from a strictly numerical point of view.  In general,
one should not expect a perfect fit to all the ``bumps and wiggles" of
the light curve because the spot model adopted is extremely
simplified. It is obvious, even from observations of the Sun, that
dark spots are neither homogeneous nor circular.  Nevertheless, the
surface features suggested by our analysis are interesting in
themselves because of the complementary and near-simultaneous evidence
reported by other investigators. 

\subsubsection{Photometric solutions including spots}
\label{secspotfits}

\begin{figure}
\figurenum{4}
\epsscale{0.80}
\plotone{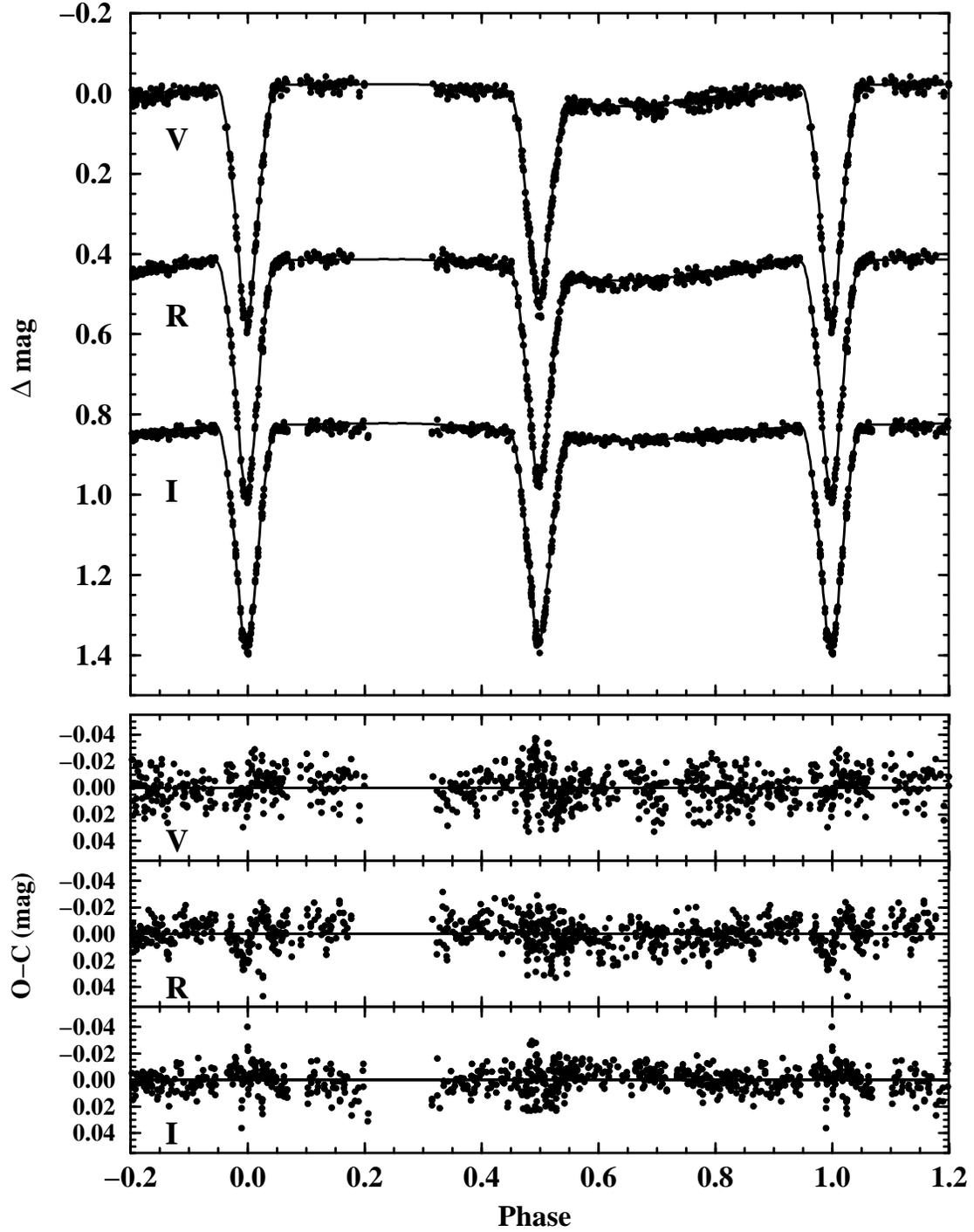}
\caption{Theoretical light curve fits to the
observations by Leung \& Schneider (1978) in the $V\!RI$ filters. The
residuals are shown at the bottom.\label{figls78curves}}
\end{figure}

For our final light curve solutions of YY~Gem we adopted the
three-spot model described above. The rms residuals from the fit were
0.013~mag, 0.011~mag, and 0.009~mag for the $V\!RI$ passbands,
respectively, in very good agreement with the individual uncertainties
quoted by \markcite{ls78}LS78.  The resulting best-fitting parameters
are listed in Table~\ref{tablightelem}.  The uncertainties given in
this table were not adopted from the (often underestimated) formal
probable errors provided by the WD code, but instead from numerical
simulations and other considerations.  Several sets of starting
parameters were tried in order to explore the full extent of the
parameter space.  In addition, the WD iterations were not stopped
after a solution was found, but the program was kept running to test
the stability of the solution and the geometry of the $\chi^2$
function near the minimum.  The scatter in the resulting parameters
from numerous additional solutions yielded estimated uncertainties
that we consider to be more realistic, and are generally several times
larger than the internal errors.  The fitted light curves are shown in
Figure~\ref{figls78curves}, with the photometric residuals at the
bottom. As mentioned earlier, both eclipses are partial and
approximately 78\% of the light of each star is blocked at the
corresponding minimum. 

As can be seen in Table~\ref{tablightelem}, the star that is slightly
more massive (photometric secondary) also appears to be slightly
cooler.  The physical significance of this result is marginal because
the components are almost identical, and in fact evolutionary
considerations would predict the opposite situation.  The most
plausible explanation for this has to do with the chromospheric
activity of the components.  Despite our best attempts to model the
main surface features of YY~Gem, in reality it may well be that the
spot filling factor is such that there are no pristine areas of the
photosphere. Large areas of unresolved spots on one star would make
the photosphere appear darker (cooler) as a whole compared to the
other star. A similar phenomenon may occur with plages.  This is
likely to change with time, one additional consequence being that the
deeper eclipse at phase 0.0 in one epoch may become shallower than the
other minimum at a different epoch. This has in fact been observed in
YY~Gem, e.g., as reported by \markcite{bdb96}Butler, Doyle \& Budding
(1996).  In performing light curve solutions such situations would
most likely be interpreted as a reversal in the surface brightness (or
effective temperature) of the components. 

Additional WD solutions were run to test the effect of other
parameters not considered above. One of these is third light ($L_3$),
which could potentially be significant due to the proximity of YY~Gem
to the bright star $\alpha$~Gem (combined $V = 1.58$, at an angular
separation of $71\arcsec$), if the background was not properly
subtracted from the photometric observations. Solutions for $L_3$ in
all three passbands were slow to converge due to the larger number of
free parameters and the inherent difficulty in fitting for third
light. The best fit was obtained for a fraction of third light of
$-1.2$\%, $-3.3$\%, and $-1.7$\%, for $V\!RI$, respectively. Negative
values can only result if the background measurements, rather than
those of YY~Gem itself, are contaminated by light from Castor, but
this appears to be ruled out by indications by \markcite{l78}LS78 that
they were very careful in making the sky readings. Furthermore, from
the early A spectral type of Castor one would expect the contribution
to be essentially ``white" light, and given the magnitudes of YY~Gem
this corresponds to fractional contributions of
1.00\th:\th0.32\th:\th0.14 for $V$, $R$, and $I$.  Thus, our solution
with $L_3$ free appears to be unphysical. Tests with fixed values of
third light of 2\%, 5\%, and 10\% in $V\!$, scaled properly to the
other bands, led to residuals from the fits that increased with
increasing $L_3$. We conclude that the best solution is obtained
without third light contribution. 

Tests were also run by including the eccentricity ($e$) and longitude
of periastron ($\omega$) among the free parameters. The resulting
values, $e = 0.010 \pm 0.005$ and $\omega = 88\fdg2 \pm 1\fdg8$, are
marginally significant. The value of $e \cos\omega$ (related to the
separation between the primary and secondary minima) is essentially
zero, indicating that the secondary eclipse lies exactly at phase 0.5
as seen earlier in \S\ref{secperiod}.  The term $e \sin\omega$,
related to the difference in width between the two eclipses, is found
to be different from zero at the 2$\sigma$ level. Rather than being
the result of an eccentric orbit, it is more likely that this is due
to imperfections in the modeling of the star spots or small
systematics in the observations themselves. The secondary eclipse
would be the most affected. We conclude that the orbit is circular
within the errors, in agreement with earlier evidence and with the
expectation from theory for a system with a period a short as that of
YY~Gem. 

Further tests were run to explore the sensitivity to changes in the
adopted mass ratio and primary temperature (upon which the
limb-darkening coefficients depend rather strongly). No perceptible
changes were seen in the results, indicating in particular that
possible inaccuracies in the Kurucz model atmosphere fluxes do not
affect the fitted light curve despite the fact that the temperature
regime at which we are operating is near the lower limit of the
validity of those models. Finally, additional WD runs with a range of
values for the bolometric albedo and the gravity-brightening
coefficient indicated that the solution cannot be improved by adopting
values different from those predicted by theory. 

\subsection{Fits to the light curve of Kron (1952)}
\label{seckron}

As mentioned earlier, the photometry of YY~Gem obtained by Kron (1952)
is available for a single passband centered near the $V$ band.
Because of the poorer quality of the measurements, their smaller
number, and the narrow wavelength coverage, the parameters from this
light curve are less reliable than those resulting from the analysis
described above. Nevertheless, a re-analysis of this material allows
us to compare with the physical properties of the stars inferred from
the \markcite{ls78}LS78 data, which we regard as an important check on
the external accuracy of the radii. In addition, it enables us to
study the activity level of the system at a different epoch. 

The WD solutions described in the previous section were used as a
starting point, with all parameter settings being identical to those
used for the \markcite{ls78}LS78 light curves except for the spot
properties and the fixed value of $\Omega_A$.  A systematic search was
carried out to find a value for $\Omega_A$ that yields a ratio of the
radii of unity when fitting the light curve.  Because of the intrinsic
scatter of the photometric measurements, the surface potentials that
lead to $k=1$ are not necessarily identical for the
\markcite{ls78}LS78 and Kron light curves. Indeed, we found that
$\Omega_A$ needed to be increased to a value of $7.48$ for the fit to
yield components of identical radii. 

\begin{figure}
\figurenum{5}
\epsscale{0.80}
\plotone{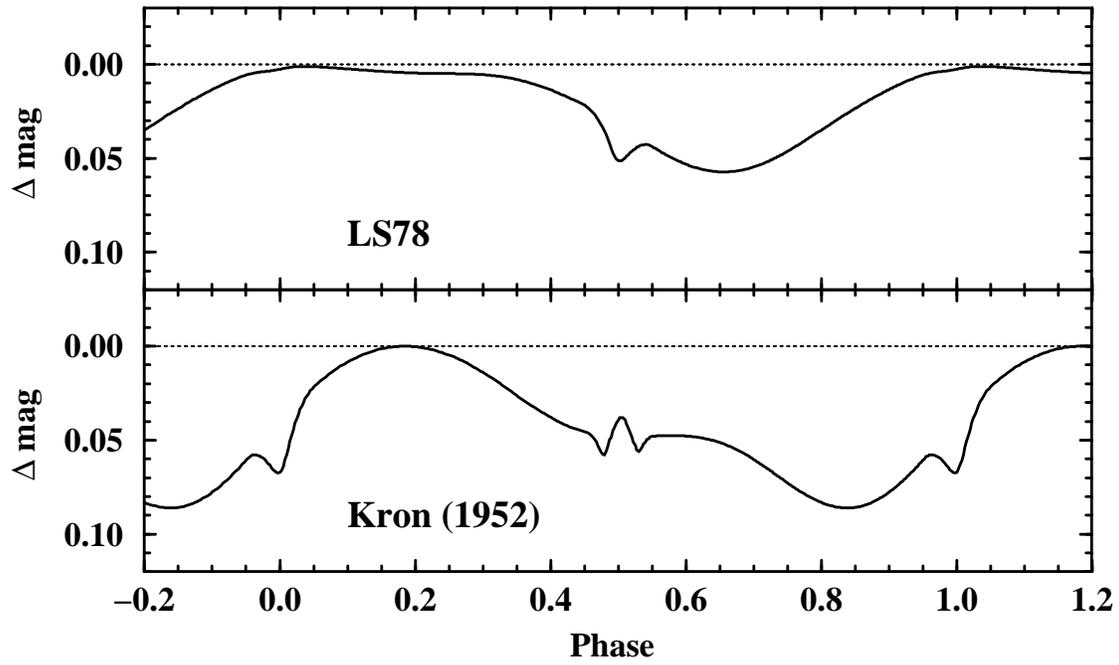}
\caption{Photometric effect of the spots on the
$V$-band light curves of LS78 and Kron (1952).\label{figspotmag}}
\end{figure}

\begin{figure}
\figurenum{6}
\epsscale{1.00}
\plotone{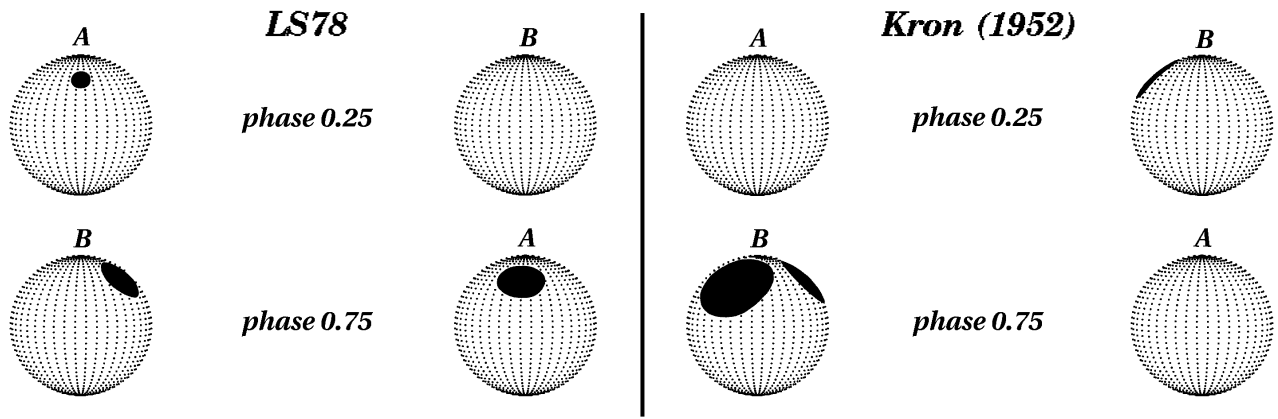}
\caption{Distribution of the modeled spots on the
surface of the stars, resulting from the light curve fits to the LS78
(left) and Kron (1952) (right) observations. The sizes and separation
of the stars are shown to scale, as they would appear to the observer
at phase 0.25 and 0.75.\label{figspotpic}}
\end{figure}

Regarding the surface inhomogeneities, the shape of the Kron light
curve suggests the presence of at least two major active regions, one
centered at the phase of the secondary eclipse and another one near
phase 0.8. WD fits considering only one dark spot proved to be
unsatisfactory. Tests were run to attempt to establish on which star
the active regions are located. The results were inconclusive,
although a marginally better fit was obtained with both spot regions
on the secondary component (as suggested also by \markcite{k52}Kron
1952 in his original analysis).  The solution we adopted has two large
spots with meridian transiting phases of 0.516 and 0.836, and radii of
$25\fdg5$ and $35\fdg0$, respectively. The sizes of these active
regions are significantly larger than those found from the analysis of
the light curves of \markcite{ls78}LS78.  This is not an unexpected
result, since the activity wave in Kron's light curve clearly has a
greater amplitude, approaching 0.1~mag. This is shown in
Figure~\ref{figspotmag}, where the photometric effect of the surface
features after subtracting out a light curve without spots is
displayed for both epochs.  Figure~\ref{figspotpic} shows the location
of the spots on the surface of the stars, for comparison with the
distribution at the epoch of the \markcite{l78}LS78 observations. 

As before, the best-fitting spot parameters presented here represent
only one of the possible configurations. Other combinations of spot
locations and sizes may yield similarly good fits, but the physical
properties of the components are essentially unaffected. 

\begin{figure}
\figurenum{7}
\epsscale{0.80}
\plotone{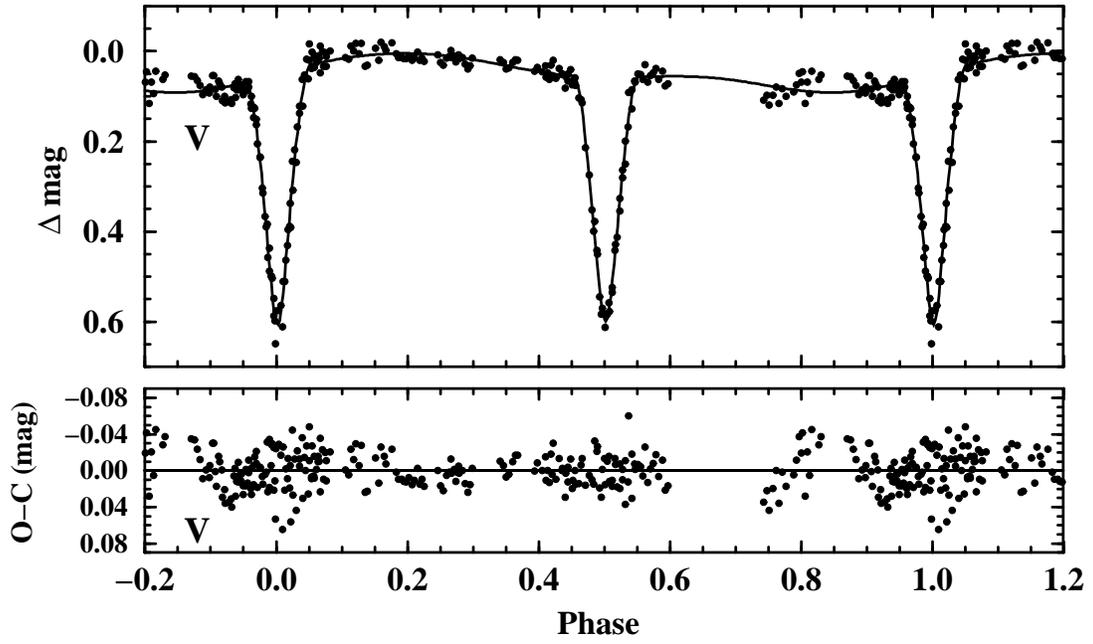}
\caption{Theoretical light curve fit to the
observations by Kron (1952) in the $V$ band. The residuals are shown
at the bottom.\label{figkroncurve}}
\end{figure}

The observations along with the theoretical light curves are shown in
Figure~\ref{figkroncurve}, and the resulting photometric parameters
from the adjustments are listed in Table~\ref{tablightelem}. The rms
residual of the fit was found to be 0.020~mag, and the $O\!-\!C$
diagram shows no significant systematic trends, indicating that the
shape of the out-of-eclipse variation has been sufficiently well
modeled (to the extent permitted by the photometric accuracy and the
model assumptions). As can be seen, the agreement between the orbital
and stellar properties derived from the analysis of the two light
curve sets (Kron and \markcite{l78}LS78) is very good. Both the
inclination angle and the mean fractional radius, the two key
geometric parameters, are found to be identical at the 1$\sigma$
level.  The only slightly discrepant property is the temperature of
the secondary star, which appears to be marginally greater than that
of the primary component from the analysis of Kron's observations.
This is an interesting result providing an illustration of our
previous argument that changes in the distribution of the spots
related to the activity can produce noticeable differences in the
apparent effective temperatures of the components. In this case, the
temperature derived for the secondary (more massive) component is
consistent with what is expected from the mass-luminosity relation. 

\section{Absolute dimensions}
\label{secabsdim}

For the final geometric properties of YY~Gem we have adopted the
weighted average of our determinations from the light curves of
\markcite{k52}Kron and \markcite{ls78}LS78. A comparison with the
original solutions by those authors reveals small differences both in
the mean radius and in the inclination angle.  Our value for the mean
relative radius ($r = 0.1589 \pm 0.0014$) is $2$\% larger than that of
Kron, and our inclination angle ($i = 86\fdg29 \pm 0\fdg10$) is about
$0\fdg1$ smaller than his\footnote{We note that \markcite{k52}Kron's
(1952) original analysis was based on a combination of his visual and
infrared data, whereas our results use only the visual observations.},
while the differences with \markcite{ls78}LS78 are $-1$\% and
$-0\fdg25$, respectively. 
		
The absolute masses and radii of the components of YY~Gem, along with
other physical parameters, follow from the spectroscopic results in
Table~\ref{tabspecelem} along with the photometric solutions in
Table~\ref{tablightelem}.  While the absolute masses we derive for the
two stars are formally different ($M_A = 0.5975 \pm 0.0047$~M$_{\sun}$
and $M_B = 0.6009 \pm 0.0047$~M$_{\sun}$), the difference is
statistically insignificant.  Similarly, the radii of the components
are so close that the observations are unable to distinguish their
ratio from a $k$ value of unity. The departure from a perfectly
spherical shape is also very small. Consistent with the equal masses
and radii, the difference in the effective temperatures (a quantity
usually best determined from the light curves) is so small that it is
completely masked by the unavoidable distortions in the photometry due
to spots, as described earlier. For all practical purposes the two
stars are therefore virtually identical in all their properties.
Accordingly, the physical quantities listed in Table~\ref{tabphys} are
the average of the formal results for the two components. 

The absolute value of the effective temperature for eclipsing binaries
is generally based on photometric calibrations external to the light
curves. For cool stars such as YY~Gem this has always been
problematic.  Few dwarfs with truly fundamental temperature
determinations are available, and many of the calibrations actually
rely on the photometric properties of giants for this temperature
regime, or even on a blackbody approximation. Others based on the
infrared flux method are model-dependent to some extent.  Significant
advances have been made in recent years in the modeling of the
photospheres of M dwarfs, used to compute synthetic colors (e.g.,
\markcite{h99}Hauschildt, Allard \& Baron 1999), and efforts to
establish a temperature scale for these stars have improved the
situation considerably (see, e.g., \markcite{k93}Kirkpatrick et al.\
1993; \markcite{lg96}Leggett et al.\ 1996; \markcite{b98}Bessell,
Castelli \& Plez 1998; \markcite{k00}Krawchuk, Dawson \& De~Robertis
2000). Nevertheless, some discrepancies still remain between
calculated colors and observed colors (see \markcite{bc98}Baraffe et
al.\ 1998).  For the present work we have relied mostly on empirical
calibrations available for a number of different color indices. YY~Gem
has been well observed photometrically from the ultraviolet to the
infrared, in a variety of systems. These measurements have been
transformed to a uniform system following \markcite{l92}Leggett
(1992), averaged, and collected in Table~\ref{tabteff}. The
calibration used for each index is also indicated in the table. In
addition, \markcite{bgh74}Bopp, Gehrz \& Hackwell (1974) extended the
infrared observations to 10~$\mu$m, and obtained another estimate of
$T_{\rm eff}$ from the spectral energy distribution (SED) of the star.
There is good agreement between all of these estimates. However, the
formal error of the average, $3820 \pm 40$~K, does not reflect
systematic errors that may be present in the calibrations. These are
rather difficult to quantify, and we have thus adopted a more
realistic value of $T_{\rm eff} = 3820 \pm 100$~K for both components.
This is the temperature we have used consistently for our
spectroscopic and photometric analyses. 

Table~\ref{tabphys} also lists the average $v \sin i$ that we measure
for the stars in YY~Gem, which is very close to the value expected for
synchronous rotation ($v_{\rm sync} \sin i$) given the very short
period of the system. We note also that the spot distortions in the
\markcite{ls78}LS78 light curves are seen to occur at the same phases
from cycle to cycle on consecutive nights, again consistent with
synchronous rotation (assuming the spots are fixed on the stellar
surface).  The luminosity and absolute bolometric magnitudes are given
in Table~\ref{tabphys} as well, based on the effective temperature and
radius. Because of the uncertainty in the bolometric corrections (BC)
for cool stars, the mean absolute visual magnitude, $M_V$, is more
accurately determined in this case from the apparent visual magnitude
of the system and the Hipparcos parallax (corrected as described in
the Appendix).  This value of $M_V$ along with the radius, in turn,
provides a check on the temperature for an adopted table of bolometric
corrections.  Using the BC scale by \markcite{f96}Flower (1996), we
obtain consistency with the radius and $M_V$ for a temperature of
3840~K and BC~$= -1.40$.  This temperature is only 20~K higher than
our mean estimate from color indices. On the other hand, if we adhere
to our temperature of 3820~K, the bolometric correction required by
the observations (radius, $M_V$) is BC~$= -1.38$, whereas the value
from \markcite{f96}Flower (1996) for this temperature is BC~$= -1.44$. 

Consistent with the poorly-determined difference in the temperatures
of the components of YY~Gem, the luminosity ratio in the visual band
is also rather uncertain. The three values for $(L_B/L_A)_V$ derived
in this paper are all different: $1.021 \pm 0.074$ (Kron light curve),
$0.886 \pm 0.010$ (\markcite{ls78}LS78 light curve), and $1.088 \pm
0.047$ (spectroscopy; \S\ref{secspecobs}). This is most likely another
manifestation of the activity of the stars, and under these
circumstances the safest assumption is that they are equally bright. 
	
\subsection{Evaluation of systematic errors in the masses}
\label{secmasses}

The paucity of good mass and radius determinations in the lower main
sequence makes it especially important to examine these estimates
critically to ensure that they are as free as possible from systematic
errors, which could compromise the comparison with theoretical models.
This is particularly true for the mass, which is the most fundamental
of all the stellar properties.  The uncertainties given above for our
mass determinations (0.8\%) are strictly internal errors. Even though
we have made every effort to minimize systematics in our radial
velocity measurements, residual errors affecting the \emph{accuracy}
are bound to remain at some level. 

\begin{figure}
\figurenum{8}
\epsscale{0.70}
\plotone{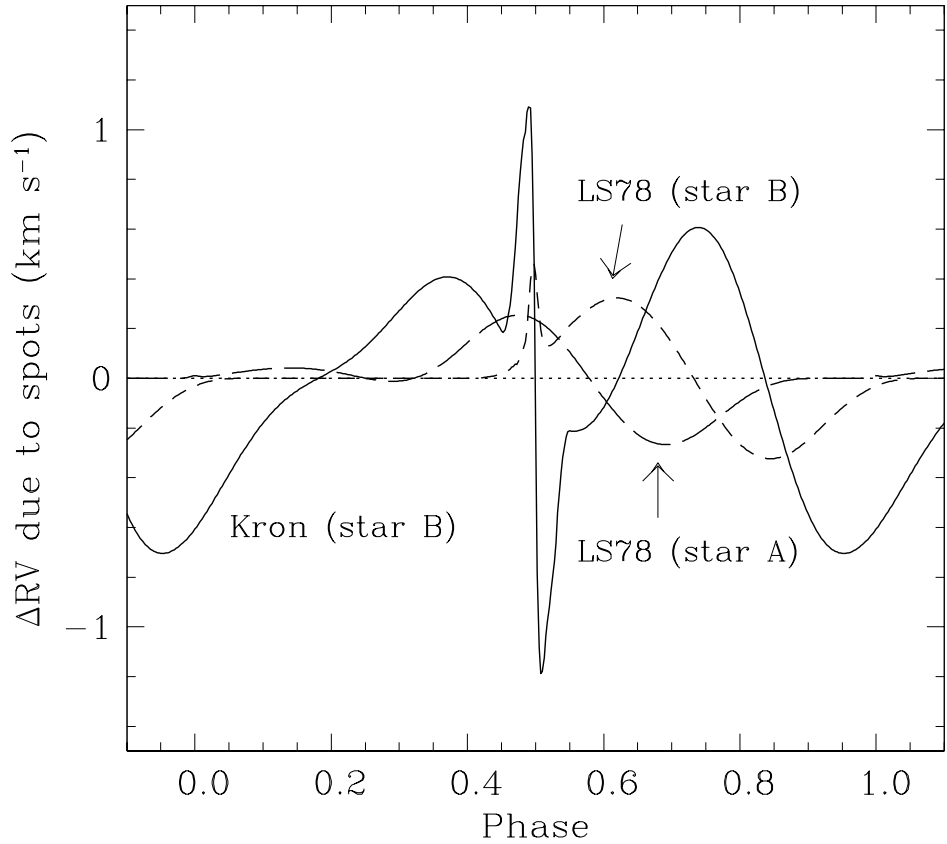}
\caption{Effect of the spots on the radial
velocities at the epoch of the LS78 and Kron (1952) observations. The
component on which the spots are located (star A or star B) is
indicated for each curve. Star A is assumed to have no spots in our
modeling of the Kron (1952) photometry.\label{figspotrvs}}
\end{figure}

One example of such errors is due to the spots present on one or both
stars. Distortions in the spectral lines produced as spots transit the
visible disk of a star can cause noticeable changes in the measured
Doppler shifts.  In Figure~\ref{figspotrvs} we show the effect that
the spots had on the radial velocities at the epoch of the photometric
observations by \markcite{k52}Kron (1952) and by \markcite{ls78}LS78,
as computed with the WD code based on the light curve fits in
\S\ref{secls78} and \S\ref{seckron}. The magnitude of the perturbation
in the velocities is relatively small, and it is unclear whether the
velocities we have measured here by cross-correlation are affected by
the full amplitude of these corrections, but presumably they could be
biased to some extent.  As a test we have applied these corrections to
our measured velocities to estimate the impact on the mass
determinations. The primary and secondary masses change by up to
0.2\%, which is smaller than our internal errors. 

The very recent study by \markcite{sg00}S00 provides a valuable
opportunity to test the external accuracy of the mass determinations
for YY~Gem, given that both studies have formal errors under 1\%.
\markcite{sg00}S00 obtained 37 spectroscopic observations of the
binary with good phase coverage, and solved for the orbital elements
by an elegant method that uses the correlation profiles directly,
instead of deriving radial velocities in the standard way, as we have
done. Those authors pointed out that their technique greatly reduces
the effective number of free parameters of the least-squares
adjustment, and results in a significant improvement in the accuracy
of the orbital elements (although, strictly speaking, we believe they
were probably referring to the \emph{precision} of the elements).
Their published masses are $M_A = 0.6028 \pm 0.0014$~M$_{\sun}$ and
$M_B = 0.6069 \pm 0.0014$~M$_{\sun}$ for the photometric primary and
secondary, respectively, which are approximately 0.9\% and 1.0\%
larger than ours.  These differences are not much larger than our
formal errors, and in principle they suggest good agreement. The
uncertainties reported by \markcite{sg00}S00, on the other hand, are
0.2\%. At this level, the variety of systematic effects that can find
their way into the mass estimates is enough of a concern that we
believe such a small error may be too optimistic for this system. 

During the refereeing process for this paper a number of issues
regarding the \markcite{sg00}S00 results emerged that are relevant for
our discussion of systematic errors in the masses. For example, we
learned that their current best fit to the observations that includes
an iterative correction for the effect of the sidelobes of the
correlation function (not accounted for in their original analysis)
gives masses that are slightly but noticeably larger than the results
they reported earlier. The nature of this effect is analogous to the
bias that we described in \S\ref{secspecobs}, and which we addressed
for our own data by using TODCOR coupled with numerical simulations.
The new values they obtain, which supersede the original results, are
$M_A = 0.6078 \pm 0.0013$~M$_{\sun}$ and $M_B = 0.6110 \pm
0.0014$~M$_{\sun}$ (Forveille 2001, priv.\ comm.). These represent an
increase of 0.8\% and 0.7\% over the published masses, or
approximately 4$\sigma$ in terms of the internal errors.  The updated
\markcite{sg00}S00 masses are 1.7\% larger than ours, a 2$\sigma$
difference. Subtle effects such as this illustrate the great care
required to avoid potentially significant systematic errors at the 1\%
level. 

Another example is the well-known Rossiter effect\footnote{This effect
is caused when the star that is in front partially occults the
approaching or receding limb of its companion. This results in a loss
of symmetry so that the Doppler shift of the star behind, when
integrated over the visible part of its disk, no longer corresponds to
its center-of-mass velocity.} (\markcite{r24}Rossiter 1924;
\markcite{m24}McLaughlin 1924), which affects the line profiles in
spectra obtained during eclipse. For YY~Gem the distortion in the
measured radial velocity can amount to 6\kms. Such observations are
best avoided, as we have done here, unless the distortions are
explicitly included in the modeling of the velocity curves. A handful
of the observations used by \markcite{sg00}S00 do indeed fall within
the eclipse phases. However, the analysis technique they use, which
bypasses the velocities altogether, is apparently immune to this
effect, at least in the case of YY~Gem (Forveille 2001, priv.\ comm.). 

As shown above, the effect of the spots on the mass determinations is
at the level of the uncertainties quoted by \markcite{sg00}S00. By
coincidence, though, their observations are virtually simultaneous
with ours so that the surface features are not likely to affect the
\emph{difference} between our mass determinations and theirs. But they
cannot be completely ignored as a potential source of systematic
errors in the absolute masses at the 0.2\% level. 

Finally, it is worth pointing out that for the highest accuracy in
mass determinations, even an issue as trivial as the numerical
constant in the classical formula for the masses of a spectroscopic
binary cannot be overlooked\footnote{The relevant quantity in this
constant is the product $G M_{\sun}$ between the Newtonian
gravitational constant ($G$) and the mass of the Sun ($M_{\sun}$),
referred to as the ``Heliocentric gravitational constant". We note
that $G$ and $M_{\sun}$ occur only as a product in the modeling of
solar system dynamics, and do not appear separately.  The value of $G
M_{\sun}$ adopted in this paper is that recommended by the IAU (e.g.,
\markcite{aa02}The Astronomical Almanac 2002; see also
\markcite{st95}Standish 1995), and leads to a numerical constant in
the formula for the masses of $1.036149\times 10^{-7}$, when the
period is expressed in days and the velocity semi-amplitudes are given
in$\kms$.}.  The difference between the value we have adopted
($1.036149\times 10^{-7}$; see footnote) and the time-honored value
$1.0385\times 10^{-7}$ (e.g., \markcite{bfm89}Batten, Fletcher \&
MacCarthy 1989) is once again at the 0.2\% level.  A slight
inconsistency we noticed between the velocity semi-amplitudes
published for YY~Gem by \markcite{sg00}S00 ($K_A$ and $K_B$ in their
Table~2) and the absolute masses they reported (Table~3) originally
led us to suspect a problem of a similar nature.  As it turns out,
however, the numerical constant they used is nearly identical to ours,
and the explanation for the inconsistency is even more
trivial\footnote{Due to a clerical error, the orbital elements
reported in Table~2 by \markcite{sg00}S00 actually correspond to an
earlier orbit that was adjusted to radial velocities, rather than to
the full correlation profiles (Forveille 2001, priv.\ comm.). The mass
values in their Table~3, on the other hand, do correspond to the final
orbit from their original analysis. The semi-amplitudes that were used
to compute the updated masses given above (including sidelobe
corrections) are $K_A = 121.87 \pm 0.12\kms$ and $K_B = 121.24 \pm
0.11\kms$.}. 

The message we wish to convey to the reader with the arguments
presented in this section is the difficulty in reaching
\emph{accuracies} (as opposed to a \emph{precision}) in the absolute
masses much better than 1\% for a system with the characteristics of
YY~Gem.  It is entirely possible that some of the astrophysical
effects mentioned above, such as spottedness, may ultimately set the
limit for stars of this class. 

\section{Discussion}
\label{secdisc}

The considerable importance of the YY~Gem system is illustrated by the
fact that, over the last three decades, it has been used on numerous
occasions together with CM~Dra to help establish the empirical
mass-luminosity (M-L) relation and the absolute calibration of the
temperature scale for M-type stars.  Because they are eclipsing (and
hence the radii are directly measurable), these two systems have
provided by far the strongest constraints on models of stellar
structure and evolution in the lower main sequence. For the most part,
recent comparisons have concluded that there is reasonably good
agreement between the models and the observations (see, e.g.,
\markcite{cb95}Chabrier \& Baraffe 1995; \markcite{lr98}Luhman \&
Rieke 1998; \markcite{ps01}Palla \& Stahler 2001). The number of model
calculations reaching down to the mass of YY~Gem that have become
available and are in current use is large enough, however, that the
situation has become rather complex.  Statements regarding the
consistency of a particular set of models with the observations are
not necessarily true for others, given that they all make different
assumptions regarding issues such as convection, the equation of
state, boundary conditions, the helium abundance, etc.  Many of these
calculations are intended specifically for the pre-main sequence
phase.  With our new and improved determinations for the mass, radius
and temperature of YY~Gem we have taken the opportunity to revisit
this in more detail. We have considered essentially all of the models
that are most often used in the current literature for chemical
compositions near solar, and we compare them not only with YY~Gem but
also against each other. 

\subsection{Comparison with stellar evolution models}
\label{secfirstcomp}

\begin{figure}
\figurenum{9}
\epsscale{0.80}
\plotone{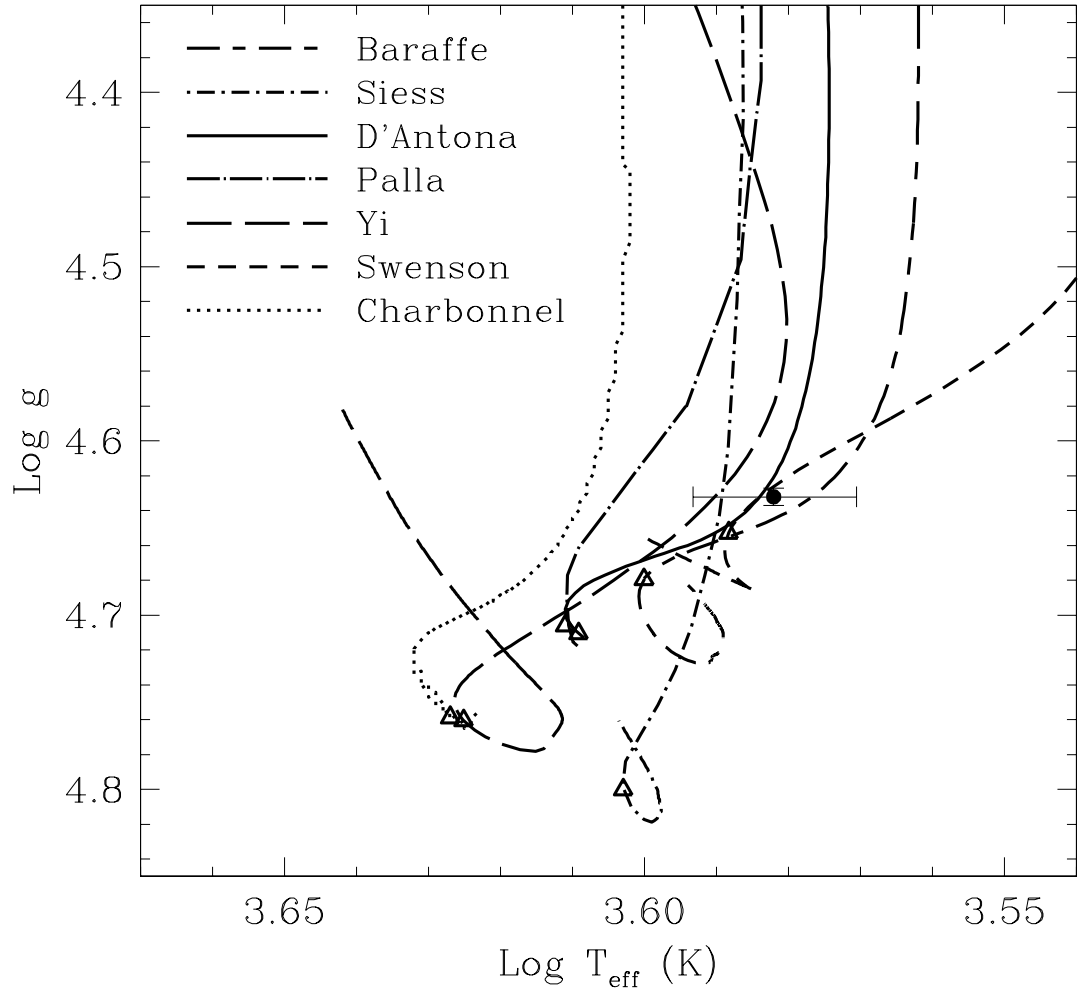}
\caption{Evolutionary tracks for solar metallicity
and a mass of 0.6~M$_{\sun}$, near that of YY~Gem, from several
different models as indicated. The triangle on each track indicates
the location of a star with an age of 100~Myr. The mean component of
YY~Gem is also shown, with the corresponding error
bar.\label{figlogg}}
\end{figure}

The mass of the mean component of YY~Gem is conveniently very nearly
equal to 0.6~M$_{\sun}$, which is one of the mass values tabulated in
all model calculations. In addition, as we show below, the metallicity
of YY~Gem is probably not far from the solar value, also the standard
reference point for abundances.  Figure~\ref{figlogg} displays
evolutionary tracks for this mass in the $\log g$ vs.\ $\log T_{\rm
eff}$ diagram from seven different model calculations for solar
composition, along with our determinations from Table~\ref{tabphys}.
The vertical section of the curves corresponds to the contraction
phase prior to the main sequence.  At first glance several of the
tracks appear consistent with the observations, within the error bars.
We note, however, that they span a broad range in effective
temperature, and there are also important displacements in the
vertical direction.  The latter have a significant impact on the age
deduced for the object from a diagram such as Figure~\ref{figlogg},
since isochrones in this plane are roughly perpendicular to the
tracks, sloping only slightly from the upper right to the lower left.
The triangles on each of the tracks indicate the locus for an age of
100~Myr, for reference. If one were to use these models and the
associated isochrones to estimate the age of YY~Gem, much in the way
this is commonly done in the H-R diagram, the results would range from
30~Myr to 85~Myr, with most of the values falling in the 30--50~Myr
range.  Laying aside for the moment the disagreement in the mass, one
would conclude that the system is very young, and that it is perhaps
still in the pre-main sequence phase.  This possibility has in fact
been advanced by \markcite{cb95}Chabrier \& Baraffe (1995), who
pointed out that a star of this mass takes approximately 300~Myr to
reach the main sequence, according to their calculations. As we show
below, however, the age of this system is almost certainly
considerably larger than the isochrones seem to indicate, which points
to a rather serious discrepancy in all the models. 

The discrepancies between the different models can be traced in most
cases to the physical assumptions, as mentioned earlier, but these
assumptions come in such a bewildering variety of combinations that it
is often difficult or impossible for the typical user to choose the
most realistic set of calculations for all applications. In recent
years the models by \markcite{bc98}Baraffe et al.\ (1998) have emerged
as one of the most commonly used in this mass regime. Compared to
other calculations, they incorporate a more sophisticated treatment of
the boundary conditions for the interior equations based on the
``NextGen" model atmospheres by \markcite{h99}Hauschildt, Allard \&
Baron (1999), which has been found to have a significant effect on the
temperature profile for cool stars. The evolutionary tracks by
\markcite{s97}Siess, Forestini \& Dougados (1997) also use a non-grey
approximation for the boundary conditions, based on analytical fits to
the model atmospheres by \markcite{p92}Plez (1992) for cool stars.
Nevertheless, even these two sets of calculations show important
differences, as seen in Figure~\ref{figlogg}, and predict ages for
YY~Gem that differ by almost a factor of two. 

The comparison in the $\log g$ vs.\ $\log T_{\rm eff}$ plane makes use
of three observable quantities: the mass, the radius, and the
effective temperature. Though important, these are insufficient for
\emph{critical} tests of stellar evolution theory in the sense defined
by \markcite{a91}Andersen (1991), i.e., tests that allow one to rule
out at least some of the calculations. This is because there are still
several free parameters left in the models that can be adjusted to
improve the fit, most notably the age and metallicity.  Furthermore,
for YY~Gem we do not have the added constraint that both components in
the binary must fit the same isochrone, because the stars are
virtually identical and reduce to a single point for all practical
purposes. 

In this particular case, however, the fact that YY~Gem is a companion
to Castor A and B, altogether forming a sextuple system, gives
additional information that we use in the next section to provide much
tighter constraints on the models.  Not only can we estimate the
metallicity of the system, but also the age, which is not normally one
of the known parameters in other binaries. Isochrone fits to Castor A
and B using a number of current models give an age of roughly 370~Myr
and a heavy element abundance of approximately $Z = 0.018$ (very
nearly the same as the solar abundance). These estimates are supported
by other analyses in the literature and an independent spectroscopic
study.  Details of these important determinations and a more complete
description of the Castor system, including the evidence that its
components are physically bound, are given in
Appendix~\ref{seccastor}. 

\subsection{A more stringent comparison with the models}
\label{secmoremodels}

With the mass, radius, temperature, absolute visual magnitude, and now
also the metallicity and age of YY~Gem known to various degrees of
accuracy, a far more demanding test of stellar evolution models can be
carried out that is usually not possible for other eclipsing binaries
in the general field and is probably unique among the low-mass stars.
In this section we have considered nine different sets of theoretical
calculations that include essentially all of the models that are most
often used or that are available for the mass regime of YY~Gem:
 \markcite{sw94}Swenson et al.\ (1994), 
\markcite{dm97}D'Antona \& Mazzitelli (1997), 
\markcite{s97}Siess et al.\ (1997) (see also \markcite{s00}Siess, Dufour \& Forestini 2000),
\markcite{bc98}Baraffe et al.\ (1998) (Lyon group), 
\markcite{ps99}Palla \& Stahler (1999),
\markcite{c99}Charbonnel et al.\ (1999) (Geneva group),
\markcite{g00}Girardi et al.\ (2000) (Padova group),
\markcite{y01}Yi et al.\ (2001) (Yale-Yonsei collaboration), and
\markcite{b01}Bergbusch \& VandenBerg (2001)\footnote{A preliminary
extension of these models to metallicities higher than [Fe/H]$= 
-0.3$ was kindly provided to us by P.\ Bergbusch (2001, priv.\
comm.).}. 
 Whenever possible we have interpolated isochrones from all of these
models for an age of 370~Myr and $Z = 0.018$, as determined from the
fits to Castor mentioned above. For the \markcite{ps99}Palla \&
Stahler (1999) models the heavy element abundance is solar and the
oldest age available is 100~Myr, but these limitations do not affect
the isochrones for YY~Gem or our conclusions significantly. 

A detailed comparison of the physical and numerical assumptions of
each set of calculations is beyond the scope of this paper. Some of
these models do not focus specifically on the lower main-sequence
although they do nominally reach small masses, and thus they may be
expected to be less successful in reproducing the observed properties
of these stars.  As mentioned earlier, the models by
\markcite{s97}Siess et al.\ (1997) and \markcite{bc98}Baraffe et al.\
(1998) implement a more sophisticated treatment of the boundary
conditions for cool stars through the use of non-grey model
atmospheres, and also use a more elaborate equation of state that
includes non-ideal effects such as pressure ionization and Coulomb
shielding.  Because of this they might be expected to give a more
realistic representation of the real properties of low-mass stars.
Thus, we have considered these two models separately in the comparison
with the observations. 

\begin{figure}
\figurenum{10}
\epsscale{0.40}
\plotone{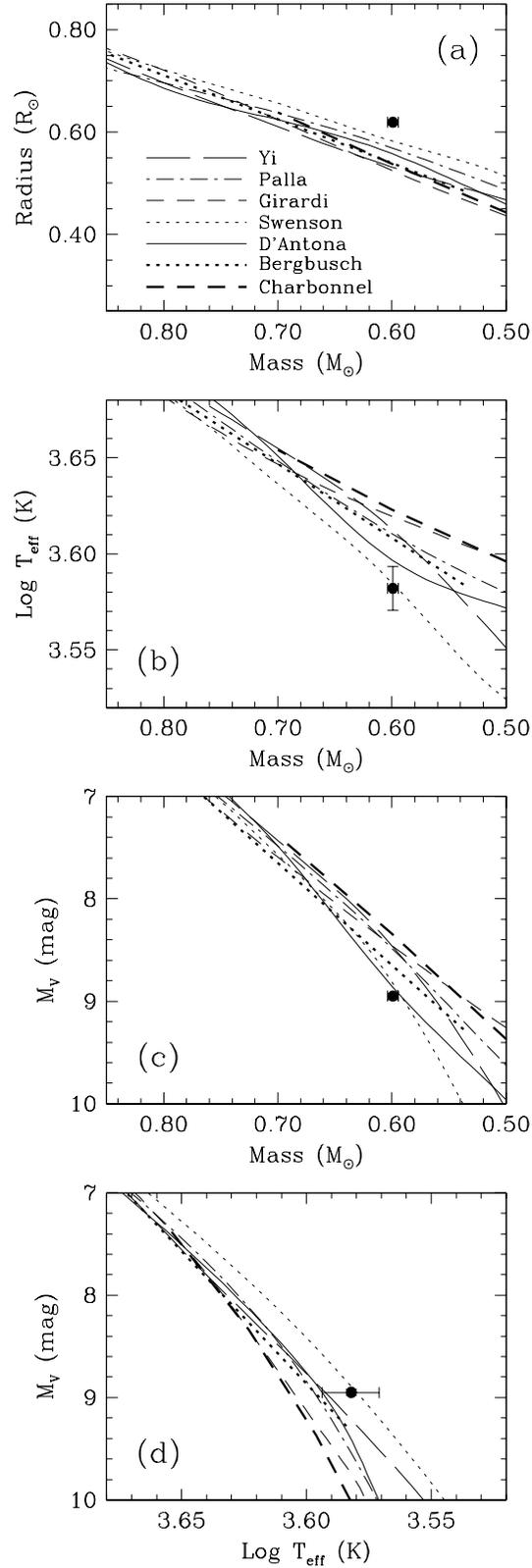}
\caption{Isochrones from seven different
theoretical models as labeled, compared to the observations of YY~Gem.
The age for the isochrones is that of Castor (370~Myr; see text), and
the metallicity adopted is $Z = 0.018$, except for the isochrones by
Palla \& Stahler (1999) that are for an age of 100~Myr and solar
abundance.\label{fig7models}}
\end{figure}

Figure~\ref{fig7models} shows this comparison on different planes, for
all low-mass models except the two mentioned above.  The mass is the
best determined of the four observable quantities ($M$, $R$, $T_{\rm
eff}$, and $M_V$), followed by the radius. These two are of a more
fundamental nature than the temperature, since they do not depend on
external calibrations.  Thus we begin by considering the mass as the
independent variable.  

The mass-radius diagram in Figure~\ref{fig7models}a clearly shows that
all models underestimate the radius, by as much as 15\% in some cases.
The Swenson model comes closer to reproducing the observed radius
($\sim$5\% difference), although the discrepancy actually corresponds
to more than 5$\sigma$. 

In Figure~\ref{fig7models}b most of the models are seen to predict
temperatures that are too hot for the mass of YY~Gem, whereas the
\markcite{sw94}Swenson et al.\ (1994) isochrone shows good agreement
with the observations. The overall spread in the predicted
temperatures in this diagram, for a mass equal to that of YY~Gem, is
nearly 200~K (5\%).  Theoretical calculations start to diverge towards
lower temperatures, where most of the assumptions become more critical
and begin to break down. 

Figure~\ref{fig7models}c shows the familiar M-L relation. The models
that fare the best are those by \markcite{dm97}D'Antona \& Mazzitelli
(1997) and \markcite{sw94}Swenson et al.\ (1994). All the others
appear to overestimate the flux in the visual band by up to 0.6~mag,
probably as a consequence of differences in the opacities and other
physical assumptions. 
	
Finally, the H-R diagram in the bottom panel suggests that all models
overestimate the temperature at the absolute magnitude of YY~Gem, with
the one exception once again of the \markcite{sw94}Swenson et al.\
(1994) isochrone, which appears to provide an excellent fit.  However,
the better agreement in this case, as in Figure~\ref{fig7models}b and
Figure~\ref{fig7models}c, is not necessarily an indication of more
sound physical assumptions. In fact, the equation of state adopted in
this model, which relies on the tabulation by
\markcite{eff73}Eggleton, Faulkner \& Flannery (1973), is rather
simple by today's standards, and, in addition, the boundary conditions
are treated within the grey approximation. In retrospect,
Figure~\ref{fig7models} shows that this model tends to stand out
compared to all the others, in all four planes. The same is seen in
Figure~\ref{figlogg}. 

Among the eclipsing binaries with well-determined component masses
under 1~M$_{\sun}$, UV~Psc (HD~7700) and V818~Tau (HD~27130) have
secondaries that are only slightly more massive than YY~Gem, and might
serve to confirm the trends seen in Figure~\ref{fig7models}. Both have
$M = 0.76$~M$_{\sun}$. UV~Psc is apparently an old and somewhat
evolved object (see \markcite{p97}Popper 1997) of unknown metallicity
and is therefore not as useful here.  On the other hand V818~Tau is a
member of the Hyades cluster, which happens to have a composition
([Fe/H]$ = +0.13$; \markcite{bf90}Boesgaard \& Friel 1990) not far from
what we have estimated for Castor. The well-known age ($\sim$600~Myr)
is somewhat older than that of YY~Gem, but this difference has a
negligible effect on the properties of stars of this mass. By
combining the light-curve solutions for V818~Tau by
\markcite{sm87}Schiller \& Milone (1987) with the spectroscopic
analysis of that system by \markcite{ps88}Peterson \& Solensky (1988)
we obtain the absolute dimensions listed in Table~\ref{tabvb22}. The
effective temperatures are based on the light curve results by
\markcite{sm87}Schiller \& Milone (1987), their derived color indices
in several passbands, and the same color-temperature calibrations used
for YY~Gem in Table~\ref{tabteff}. For the absolute visual magnitude
we have adopted the Hipparcos parallax of the star. 

\begin{figure}
\figurenum{11}
\epsscale{0.40}
\plotone{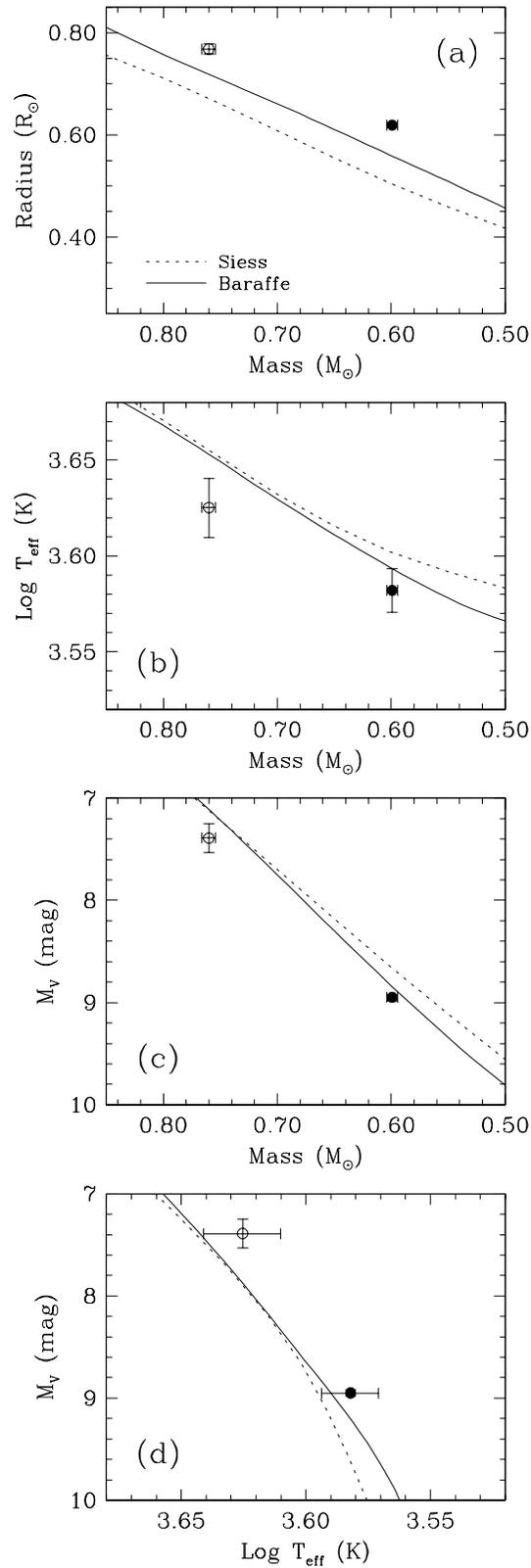}
\caption{Same as Figure~\ref{fig7models}, for the
models by Siess et al.\ (1997) and Baraffe et al.\ (1998) and an age
of 370~Myr and $Z = 0.018$. In addition to YY~Gem (filled circle), we
show also for comparison the secondary component of another eclipsing
binary in this mass regime (V818~Tau, in the Hyades cluster),
represented with an open circle. The error bars in the top panel are
the same size as the symbols.\label{fig2models}}
\end{figure}

In Figure~\ref{fig2models} we show the observations for YY~Gem and
also V818~Tau~B, along with the isochrones by \markcite{s97}Siess et
al.\ (1997) and by \markcite{bc98}Baraffe et al.\ (1998).  Models for
the exact composition of V818~Tau~B, which is slightly higher than
solar, show only a very small difference in these diagrams, and have
no effect on the conclusions. 

Figure~\ref{fig2models}a indicates, as before, that the predicted
radii are too small, by 10\% for \markcite{bc98}Baraffe et al.\ (1998)
and nearly 20\% for \markcite{s97}Siess et al.\ (1997). This
represents more than 10 and 20 times the uncertainty in the measure
radii, and is therefore highly significant. The explanation for the
age discrepancy pointed out in reference to Figure~\ref{figlogg} is
now obvious: YY~Gem appears too large compared to predictions, and the
models therefore require it to be at an earlier stage of evolution,
still contracting towards the main sequence\footnote{A much later
stage of evolution (post-main-sequence) is also able to match the
observed radius, but only at an age of several tens of Gyr that is
totally inconsistent with the main-sequence status of Castor A and B.
Also, the predicted temperature for YY~Gem would be some 500~K hotter
than observed.}. 

The agreement between the two isochrones themselves in the $\log
T_{\rm eff}$-mass plane (Figure~\ref{fig2models}b), the
mass-luminosity diagram (Figure~\ref{fig2models}c), and the H-R
diagram (Figure~\ref{fig2models}d) is moderately good except at the
low-mass end.  From the first and third of these, however, it would
appear that, although the slope of the models seems correct, both
isochrones tend to overestimate the effective temperature by roughly
150~K at a constant mass or a constant $M_V$, if the empirical
color-temperature calibrations are to be relied upon.  For each star
the discrepancy is admittedly only at the level of 1-1.5$\sigma$, but
the fact that it is in the same direction for both is suggestive, and
may become more significant if additional systems in this mass regime
confirm the trend.  The modest match between the observations and the
theoretical mass-luminosity relation in Figure~\ref{fig2models}c is
similar to that found in the visual band by \markcite{d00}Delfosse et
al.\ (2000) from a larger sample of astrometric binaries, and may
indicate that the model luminosities are somewhat overestimated, as
mentioned earlier. Those authors also showed that the agreement in the
$J\!H\!K$ passbands tends to be better. 

The discrepancy in the radii for main-sequence stars below
1~M$_{\sun}$ has occasionally been pointed out before in the
literature (e.g., \markcite{cb95}Chabrier \& Baraffe 1995;
\markcite{l99}Lastennet et al.\ 1999) but its importance has often
been downplayed because of the additional degrees of freedom in the
models (mainly age and metallicity) or the low precision of the
observations.  Both of these shortcomings are removed in the present
paper so that the disagreement is now more obvious.  The problem was
expressed in terms of ages by \markcite{p97}Popper (1997), who noticed
that the secondary stars in eclipsing binaries in the
0.7-1.1~M$_{\sun}$ range with components differing appreciably in mass
and radius appear systematically older than their primaries by factors
of two or more.  \markcite{cbcv99}Clausen et al.\ (1999) added more
data and reinterpreted the deviations in terms of the radii, as we
have done here. 

This result emphasizes the importance of having the radii as a
sensitive indicator of evolution, and the need for the stellar
evolution models to satisfy \emph{all} observational constraints
\emph{simultaneously} in order to pass the test.  As we see in the
system discussed here, the temperature and radius discrepancies tend
to cancel each other out to some degree and lead to luminosities that
are not too far off, at least compared to some models. So, for
example, if only the mass-luminosity diagram is considered
(Figure~\ref{fig2models}c, or Fig.~3 by \markcite{d00}Delfosse et al.\
2000), one might be led to believe that the agreement is perhaps
satisfactory within the errors and that no action is required on the
theoretical side to improve the fit.  Unfortunately this is very often
the situation for the lower main-sequence, largely because of a lack
of other observational data (particularly radius or surface gravity
determinations).  The perception in the community that all is well
because the fit to the M-L relation is tolerably good, within
observational errors, is not uncommon.  In YY~Gem the discriminating
power of the mass, radius, and temperature determinations is further
enhanced by the fact that the metallicity and even the age can be
estimated independently, which makes the test much more demanding. A
common practice in the field of star formation to estimate absolute
ages and masses for T~Tauri stars is to place the objects on the H-R
diagram, if the distance is known. In the particular case of YY~Gem or
V818~Tau this procedure leads to masses that are not far off from
their true values because the evolutionary tracks happen to be mostly
horizontal so that the temperature errors have little effect.  The
ages, however, are still underestimated by factors of 5 and 10,
respectively, which highlights the limitations of this popular
analysis tool for low-mass stars. 

Another issue that we view as equally important is the need to test
more than one set of models, given the fact that the assumptions are
so varied that all models are different. This is the motivation for
our efforts to collect and compare a large number of the most often
used theoretical calculations from the recent literature (nine) for
the mass regime of YY~Gem.  Figures \ref{fig7models} and
\ref{fig2models} show that \emph{all} models fail this very strict
comparison with YY~Gem, even the ones considered to be more realistic.
Furthermore, this conclusion is supported by a second object of nearly
the same metallicity (V818~Tau) with similarly good determinations of
$M$, $R$, $T_{\rm eff}$ and $M_V$. Thus, it is clear that the physics
of low-mass stars is not as well understood as one may gather from
some recent claims. 

The reason for the failure of the models to reproduce the radii (and
temperature) of stars of this mass is not easy to pinpoint. One
obvious candidate is the treatment of convection. Most models use the
standard mixing-length prescription (\markcite{bv58}B\"ohm-Vitense
1958) with a mixing length parameter $\alpha$ in the range 1.5-2.0,
close to that required for the Sun. As pointed out by
\markcite{cb95}Chabrier \& Baraffe (1995) this parameter can have a
significant effect in stars like YY~Gem, which develop radiative cores
extending more than 70\% in radius.  Compared to other calculations
their models use a considerably lower value of $\alpha = 1.0$, yet an
even smaller value would be required to fit the observed radius, which
they view as unrealistic.  In addition, much lower values of $\alpha$
appear to disagree with results from recent hydrodynamical
calculations (\markcite{f99}Freytag, Ludwig \& Steffen 1999).  An
alternative scheme for convection is the ``Full Spectrum of
Turbulence" prescription (\markcite{cm91}Canuto \& Mazzitelli 1991).
Of all the models tested here it has been implemented only in the
\markcite{dm97}D'Antona \& Mazzitelli (1997) calculations, and tends
to predict higher temperatures than the standard theory. The equation
of state, which largely determines the mechanical structure of stars
in the low-mass regime, may also be in need of improvement.  Very
recent work by \markcite{mm01}Mullan \& MacDonald (2000) even explores
the effect of magnetic fields on the structure of low-mass stars.
Their models show that calculations that \emph{do not} incorporate
magnetic fields tend to predict radii that are too small and effective
temperatures that are too high, exactly as seen above for YY~Gem.
Though this finding is very encouraging, most likely the observed
discrepancies with the observations in this mass range will turn out
to be due to a combination of several model ingredients. A careful
investigation of lower-mass systems such as CM~Dra would probably shed
some light on these issues, since convection becomes less important
for smaller stars while the denser structure and cooler temperatures
should highlight possible deficiencies in the equation of state. 
	
The interesting question of whether or not YY~Gem has already reached
the Zero Age Main Sequence (ZAMS), raised earlier, is more complicated
to answer than would appear. According to \markcite{cb95}Chabrier \&
Baraffe (1995) a 0.6~M$_{\sun}$ star of solar metallicity arrives on
the main sequence at an age of $\sim$300~Myr.  \markcite{ps99}Palla \&
Stahler (1999), on the other hand, give an age of only 85~Myr. The
ZAMS computed by \markcite{s00}Siess et al.\ (2000) that does not
include convective core overshooting suggests that the age for this
mass is much older (600~Myr). When overshooting is accounted for,
their models give an even older age of 1~Gyr. A closer investigation
reveals that the definitions adopted for the ZAMS by each group, which
are arbitrary to some extent, are not the same.  For example,
\markcite{ps99}Palla \& Stahler (1999) defined the ZAMS to be the
point where the luminosity released through gravitational contraction
falls to 3\% of the total. \markcite{s00}Siess et al.\ (2000) define
it as the point when 0.1\% of the central hydrogen mass fraction has
been burnt (Siess 2001, priv.\ comm.). If a definition similar to that
by Palla \& Stahler based on the fraction of gravitational energy is
adopted for the Siess models that include overshooting, the result is
$\sim$100~Myr. The ZAMS is thus not a very clearly defined concept, or
at least there seems to be no consensus on a practical definition.
This, along with the discrepancies noted above in the comparison with
the observations, makes it virtually impossible to decide on the exact
evolutionary status of YY~Gem. 
	
\section{Final remarks}
\label{secremarks}

The lower main sequence is one of the areas of the H-R diagram where
stellar evolution models are least constrained by the observations.
YY~Gem (along with CM~Dra) remains one of the key systems in this mass
regime. Our new spectroscopic observations have yielded masses for the
virtually identical components good to better than 1\%, differing by
2$\sigma$ (1.7\%) from another recent determination
by \markcite{sg00}S00 (updated as reported in \S\ref{secmasses})
with formal errors also under 1\%.  Our emphasis in this paper has
been on understanding the systematic errors, perhaps the most
insidious component of the total uncertainty at this level of
precision. 

A re-analysis of existing light curves with more sophisticated
techniques than available at the time of the original studies has
allowed us to derive the radius of the mean component to a precision
also better than 1\%, nearly a factor of 4 improvement over the value
listed in the compilation by \markcite{a91}Andersen (1991). The radius
is one of the crucial stellar properties because of its sensitivity to
evolution. The difference with the original determinations is 1-2\%,
but the errors are now much better characterized. 

To strengthen the comparison with the models we have made use of the
fact that YY~Gem is a member of the Castor sextuple system. The
assumption of a common origin, and therefore of a common age
($\sim$370~Myr) and metal abundance ($Z = 0.018$) provides important
constraints that are not normally available in other eclipsing
binaries.  The observations for YY~Gem show quite convincingly that
all current stellar evolution models underestimate the radius of stars
of this mass by up to 20\%, and also that they tend to overestimate
the temperature.  A remarkably similar trend seen for V818~Tau,
another well-observed eclipsing binary in which the secondary is only
slightly more massive, makes these results even more compelling.  As a
direct consequence of the radius problem, ages for stars of these
characteristics are underestimated by factors of 4 to 10, depending on
the model. 

This discrepancy is a matter of some concern, given that many of these
models are commonly used to estimate absolute ages for young open
clusters, or even to date individual T~Tauri stars in the H-R diagram.
It is possible that offsets in temperature and radii cancel each other
out to some extent in the classical H-R diagram or in the mass-luminosity
diagram for certain mass regimes, but we believe that such ages must
still be viewed with suspicion. It may well be that for ages much
younger than that of YY~Gem the deviations are not as important, but
until this is shown to be the case, caution is advised. 

The theory of stellar evolution in the lower main-sequence has made
great strides in the last decade or so, and our current state of
knowledge represents quite a significant achievement given the
complexity of the physics.  But it is clear that adjustments are still
needed, and despite the impression one may receive from some studies
in the recent literature that focus only on the H-R diagram or the M-L
diagram, the ability to reproduce all of the observations
simultaneously is nowhere near that of the models for higher-mass
stars, where suitable eclipsing binaries are much more numerous and
provide greater constraints. In fact, our use of the latter models to
set the age and metallicity of Castor gives the tests described in
this paper for YY~Gem a differential character, in a sense. To the
extent that we trust the higher-mass models, the comparison with the
observations highlights the deficiencies in the theory of low-mass
stars. 

Favorable systems for critical tests of the models are unfortunately
still very few in this mass regime, but thanks to the efforts by
\markcite{p96}Popper (1996) and others, good progress is being made in
the identification of additional candidates (see, e.g.,
\markcite{cbcv99}Clausen et al.\ 1999). Follow-up spectroscopic and
photometric studies of these objects should contribute greatly to the
advancement in this field. 
	
\acknowledgments

Thanks are due to P.\ Berlind, J.\ Caruso, D.\ W.\ Latham, A.\ Milone,
R. P.\ Stefanik, and J.\ Zajac, who obtained many of the spectroscopic
observations, and to R.\ J.\ Davis, who maintains the CfA database of
radial velocities. Helpful comments and suggestions were provided by
the referee, Dr.\ T.\ Forveille, including unpublished information
regarding the \markcite{s00}S00 results.  We are also grateful to L.\
Siess, F.\ Palla, S.\ Yi, C.\ Charbonnel, and P.\ A.\ Bergbusch for
helpful comments and assistance with their evolutionary model
calculations, and to M.\ Standish and B.\ Marsden for discussions
regarding fundamental constants. I.\ R.\ acknowledges financial
support from the Catalan Regional Government (CIRIT) through a
Fulbright fellowship.  This research has made use of the SIMBAD
database, operated at CDS, Strasbourg, France, and of NASA's
Astrophysics Data System Abstract Service. 

\appendix
\section{Appendix: Castor as a sextuple system}\label{seccastor}

We describe here the details of the determination of the age and the
heavy element abundance of Castor. These are two of the key properties
that we use in \S\ref{secmoremodels} to further constrain stellar
evolution models for YY~Gem, by virtue of the physical association
(also discussed below) of the eclipsing system with the two brighter
stars. 
	
As mentioned in \S\ref{secintro}, both Castor A and Castor B are
single-lined spectroscopic binaries so that the secondary components
do not contribute any significant light and the photometry of each
system corresponds to that of the primaries.  Their position in the
H-R diagram ($L/L_{\sun}$ or $M_V$, and $T_{\rm eff}$) may therefore
be used to estimate their physical properties by comparison with
theoretical calculations. 

The Castor AB system was observed and resolved by the ESA astrometry
mission Hipparcos. The Double and Multiple Systems Annex to the
Hipparcos catalog (\markcite{esa}ESA 1997) lists $H_p$ magnitudes of
$1.934\pm0.004$ and $2.972\pm0.009$ for Castor A and B, respectively,
along with a trigonometric parallax of $\pi_{\rm HIP} = 63.27 \pm
1.23$~mas. These are the main ingredients needed to compute the
luminosities of the stars.  In deriving these parameters Castor~A and
Castor~B were assumed to have the same parallax (a valid assumption),
but they were also assumed to have a common proper motion, which is
not necessarily true due to the orbital motion of the pair. In fact,
according to the elements by \markcite{h88}Heintz (1988) the relative
motion over the 3-year interval in which the object was observed by
Hipparcos is 345~mas, a non-negligible amount given the typical
precision of $\sim$1~mas for the satellite measurements. The parallax
is thus suspect, and the magnitudes may be affected as well.  To
remedy this situation we have re-reduced the Hipparcos transit data
accounting for the orbital motion of the visual binary.  The relative
motion predicted by Heintz's orbit over the interval of observation is
essentially linear: the deviation from a linear trajectory is only
1~mas in Right Ascension and 0.5~mas in Declination. For all practical
purposes, therefore, it is sufficient to allow the two stars to have a
different proper motion in the new solution. The parallax resulting
from this new fit is $\pi_{\rm HIP} = 66.90 \pm 0.63$~mas, with an
error only half as large as before. The difference with the old value
represents a change of 2.6$\sigma$ (3.6~mas). We discuss the
implication of this below. 

The $H_p$ magnitudes also change slightly in our new fit. The new
values are $1.933\pm0.001$ and $2.978\pm0.004$ for Castor A and B,
respectively.  We transformed these to standard Johnson $V$ magnitudes
by making use of Table~1.3.5 of the Hipparcos catalog Volume~1
(Introduction and Guide to the Data). This table provides differences
between $H_p$ and $V$ as a function of $(V\!-I)_{\rm C}$ (in the
Cousins system).  Unfortunately, no individual measurements of this
color index are available for the components of Castor. To circumvent
the problem, we used the spectral type of the components as a first
approximation and then checked for consistency with the final set of
temperatures and color indices.  The most recent and thorough spectral
classification of the individual components of Castor is that of
\markcite{e76}Edwards (1976). He classified Castor~A as A1~V and
Castor~B as A5~Vm (note that the combined system was assigned a
spectral type of A2~Vm, which has been a source of some confusion in
later references and even in the SIMBAD database). Following the
calibration by \markcite{b79}Bessell (1979), these spectral types
translate into $(V\!-I)_{\rm C} = +0.015$ and $(V\!-I)_{\rm C} =
+0.155$ for Castor A and B, respectively.  The transformed $V$
magnitudes with estimated uncertainties are therefore $V_A =
1.93\pm0.02$ and $V_B = 2.93\pm0.02$. 

The trigonometric parallax of Castor has been measured from the ground
numerous times prior to the Hipparcos mission.  The most recent
compilation of ground-based values (Yale Catalogue; \markcite{va95}van
Altena, Lee \& Hoffleit 1995) lists 16 entries, but does not
distinguish which are for Castor itself and which are for YY~Gem, and
simply averages all determinations.  The matter of the physical
association between the eclipsing binary and the two bright stars is
of considerable interest here because of our assumption that they have
a common physical origin. As it turns out, this is actually
corroborated by other evidence, as described in more detail below.
Referring back to the original sources for the trigonometric
parallaxes we find that 6 of the measurements are for YY~Gem, and the
remainder are for Castor A, Castor B, or their average. The weighted
means for YY~Gem and for Castor are statistically indistinguishable
from each other, as well as from the average of all 16 measurements
(which is $\pi_{\rm trig} = 74.8 \pm 2.5$~mas).  This, however, is
significantly larger (by 3$\sigma$) than the value we determined above
based on a re-analysis of the Hipparcos transit data ($\pi_{\rm HIP} =
66.90 \pm 0.63$~mas).  There is a rather large scatter among the
individual measurements from the Yale Catalogue, which in the case of
Castor is to be expected due to the brightness and close proximity of
the visual pair.  Recent studies of YY~Gem by
\markcite{sg00}S\'egransan et al.\ (2000) and \markcite{d00}Delfosse
et al.\ (2000) have adopted the mean parallax from the Yale Catalogue.
We have preferred to rely here on the more precise determination based
on the Hipparcos mission, a choice that translates into a
non-negligible difference in the distance modulus of $\Delta(m\!-\!M)
= 0.24$~mag, making the stars intrinsically brighter.  With this, the
absolute magnitudes for Castor A and B are $M_V = 1.05 \pm 0.02$ and
$M_V = 2.05 \pm 0.02$, respectively, in excellent agreement with the
expected absolute magnitudes of dwarf stars of these spectral types
(see \markcite{sk82}Schmidt-Kaler 1982).  Using bolometric corrections
taken from the calibration by \markcite{f96}Flower (1996) the absolute
luminosities are $\log L_A/L_{\sun} = 1.52 \pm 0.01$ and $\log
L_B/L_{\sun} = 1.06 \pm 0.01$. 

An estimate of the effective temperatures can be obtained from the
spectral type--luminosity class--temperature calibrations of
\markcite{jn87}de Jager \& Nieuwenhuijzen (1987). This yields values
of $T_{\rm eff} = 9400$~K and $T_{\rm eff} = 8180$~K for Castor A and
B, respectively.  An independent determination based on color indices
would be valuable, but again, no individual color indices for the
components are available in the literature. Only combined values of
$(B\!-\!V) = +0.04 \pm 0.01$ and $(U\!-\!B) = +0.02 \pm 0.01$ have
been published by \markcite{n78}Nicolet (1978), and by several earlier
sources giving nearly identical results.  Individual indices can be
derived from the joint photometry if the luminosity ratio between the
components at each passband is known. The luminosity ratio in the $V$
band, $q_V$, follows directly from the difference in the apparent
magnitudes: $q_V = 0.398$ (secondary/primary). To estimate the
luminosity ratios in the $B$ and $U$ passbands we made use of
synthetic photometry computed from Kurucz ATLAS9 models normalized
according to the luminosity ratio in $V$, and adopting also effective
temperatures of 9500~K and 8250~K, solar metallicity, and $\log g =
4.0$. Tests indicated a negligible dependence of the computed
luminosity ratios on the adopted metallicity and $\log g$.  The ratios
obtained were $q_B = 0.353$ and $q_U = 0.324$, which lead to the
following individual colors:

\[\begin{array}{ll}
(B\!-\!V)_{\rm A} = 0.00,~~~ & (B\!-\!V)_{\rm B} = +0.13 \\
(U\!-\!B)_{\rm A} = 0.00,~~~ & (U\!-\!B)_{\rm B} = +0.09
\end{array}\]

Effective temperatures for the components from the Johnson broad-band
color indices were derived using several empirical calibrations:
\markcite{p80}Popper (1980), \markcite{bv81}B\"ohm-Vitense (1981),
\markcite{sk82}Schmidt-Kaler (1982), \markcite{f96}Flower (1996), and
\markcite{b98}Bessell et al.\ (1998).  An average of all
determinations gives values of $T_{\rm eff} = 9420 \pm 100$~K and
$T_{\rm eff} = 8240 \pm 150$~K for Castor A and B. These temperatures
are in good agreement with the preliminary estimates from the spectral
types, which ensures that the analysis described herein is fully
self-consistent. 

The metallicity and age of Castor may now be estimated by comparison
with theoretical isochrones, a number of which are available for this
mass regime. Isochrone fitting based on an observed luminosity and
temperature for a given star, or even for \emph{two} stars as in this
case, does not generally give a unique answer.  There are still
several free parameters in the models that can be adjusted to produce
equally good fits, such as the mixing length parameter, convective
core overshooting, the helium abundance, etc. In practice, however,
these parameters are fixed by the authors of the models to some
particular value deemed to be reasonable, and so the result of the fit
to the observations will depend to some extent on those assumptions.
It will also depend on the particular calibration chosen for the
conversion from color indices to effective temperatures, typically
built into the published isochrone tables, and to a lesser degree on
the scale of the bolometric corrections. To provide some sense for the
range of ages and metallicity estimates ($Z$) that one can obtain for
Castor from isochrone fitting, Table~\ref{tabagemet} collects our
determinations based on five of the most commonly used models. 

For several of them we list more than one determination, which
illustrates the effect of a difference in the bolometric corrections
adopted (``BC" heading in column~1). For the color/temperature
conversion we rely on an average of various calibrations, as described
above, and thus we compare theoretical and observational temperatures
directly, without making use of the calibrations built into the
isochrones. The effect of a difference in helium abundance ($Y$) is
tested using the models by \markcite{cg98}Claret \& Gim\'enez (1998),
which allow a choice for this parameter (in all other models $Y$ is
set by the enrichment law adopted). A lower helium abundance leads to
an older age and a lower metal abundance. 

The ages for the first four models in Table~\ref{tabagemet} range from
333~Myr to 410~Myr, a representative value being 370~Myr. The models
by \markcite{s97}Siess et al.\ (1997) consistently give somewhat
younger ages (along with larger masses) because they do not account
for convective core overshooting, which has the effect of prolonging
the hydrogen-burning phase.  A slightly lower value for the age ($200
\pm 100$~Myr) was derived in a study by \markcite{bn98}Barrado y
Navascu\'es (1998) on the basis of the kinematics of Castor as well as
lithium abundances, isochrone fitting, and activity indicators for a
number of putative members of the Castor moving group. 

The metallicities in Table~\ref{tabagemet} that best fit Castor are
all seen to be quite near the solar value, for a range of values of
$Y$. We adopt a representative abundance given by $Z = 0.018$. Direct
observational support for this is scarce.  Despite its brightness,
Castor has rarely been studied spectroscopically at high dispersion to
determine the abundance because of the close separation of the visual
pair. The work by \markcite{s74}Smith (1974) appears to be the only
source available, and gives [Fe/H]~$= +0.7$ (average for the two
stars) relative to Vega, although with a large error.  Twenty-one
independent determinations of the metallicity of Vega are listed in
the latest edition of the catalogue of [Fe/H] determinations by
\markcite{cs97}Cayrel de Strobel et al.\ (1997). The scatter is quite
large, but the more recent values cluster around [Fe/H]~$= -0.6$. The
observed metallicity of Castor relative to the Sun is therefore
[Fe/H]~$= +0.1 \pm 0.2$, consistent with the results from our
isochrone fitting. 

\begin{figure}
\figurenum{12}
\epsscale{0.80}
\plotone{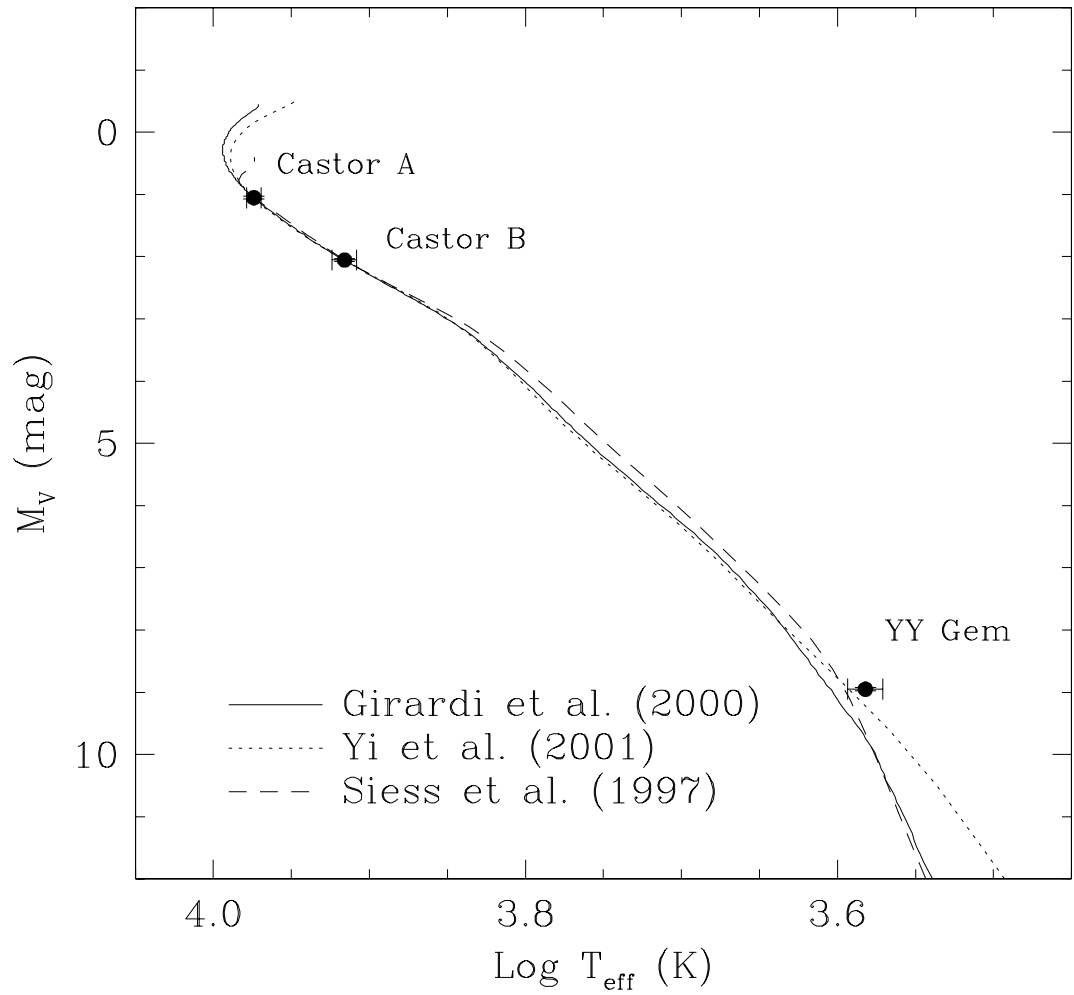}
\caption{Isochrone fits to Castor A and B using
three different models. The ages for these isochrones are 333~Myr
(Girardi et al.\ 2000), 378~Myr (Yi et al.\ 2001), and 300~Myr (Siess
et al.\ 1997; no overshooting), and the chemical compositions are
listed in Table~\ref{tabagemet}.  YY~Gem is also shown, and lies
slightly above the three isochrones.\label{figcastor}}
\end{figure}

Table~\ref{tabagemet} includes also the mass and radius inferred for
the visible components of Castor A and B (referred to as ``Aa" and
``Ba") from the best-fitting isochrones. Three of the isochrones that
also reach the lower main-sequence are shown in
Figure~\ref{figcastor}, and agree by construction with the main
components of Castor but show some differences for cooler
temperatures. The observational point for YY~Gem is also shown. 

The physical association between Castor AB and YY~Gem has been studied
by \markcite{a89}Anosova, Orlov \& Chernyshev (1989) and
\markcite{ao91}Anosova \& Orlov (1991). They used all available
kinematic information (positions, proper motions, radial velocities,
and distances), and concluded that the total energy of the visual
triple is negative, so that Castor A, B, and C are gravitationally
bound to each other and most likely share a common origin.  It is
reasonable to assume, therefore, a common age and metallicity, which
removes these degrees of freedom from the comparison between the
observations for YY~Gem and the models and provides for a much
stronger test of theory. 

It is interesting to note that the masses of all six components of the
Castor multiple system can be inferred. Those of the two brighter
stars (for which we adopt averages of $M_{\rm Aa} = 2.27$~M$_{\sun}$
and $M_{\rm Ba} = 1.79$~M$_{\sun}$; Table~\ref{tabagemet}) depend upon
stellar evolution models.  From the orbital properties of the AB pair
($a_{\rm AB}=6\farcs805$, $P_{\rm AB}=467$~yr; \markcite{h88}Heintz
1988) combined with the revised Hipparcos parallax, the total mass of
this system is 4.83~M$_{\sun}$. The fractional mass ($M_B/(M_A+M_B) =
0.436$) by Heintz then leads to total masses for Castor A and B of
2.72~M$_{\sun}$ and 2.11~M$_{\sun}$, respectively. By subtraction, we
obtain the masses of the unseen companions of the two bright stars
(albeit with a rather large uncertainty) as $M_{\rm Ab} =
0.45$~M$_{\sun}$ and $M_{\rm Bb} = 0.32$~M$_{\sun}$.  These are
therefore also model-dependent.  Because the spectroscopic orbits for
the two close single-lined binaries are known, their mass functions
($f_A(M)=0.001334$~M$_{\sun}$ and $f_B(M)=0.009831$~M$_{\sun}$,
respectively; \markcite{vh40}Vinter-Hansen 1940) allow one to estimate
roughly what the inclination angles of both orbits are. A summary of
the physical and orbital properties of the sextuple system is given in
Table~\ref{tabtriple}.  Finally, the apparent separation between
Castor AB and YY~Gem translates into a projected linear separation of
about 1060~AU, which implies a minimum orbital period of approximately
$14,\!000$~yr for the observed total mass of the sextuple system of
$6.03$~M$_{\sun}$.

\newpage 

\begin{deluxetable}{ccrrrr}
\tablenum{1}
\tablewidth{30pc}
\small
\tablecaption{Radial velocity measurements and residuals for YY~Gem.\label{tabrvs}}
\tablehead{
\colhead{HJD} & \colhead{} & \colhead{RV$_{\rm A}$} & \colhead{RV$_{\rm B}$} &
\colhead{(O-C)$_{\rm A}$}  & \colhead{(O-C)$_{\rm B}$} \\
\colhead{\hbox{~~(2,400,000$+$)~~}} & \colhead{Phase\tablenotemark{a}} & \colhead{(\kms)} &
\colhead{(\kms)} & \colhead{(\kms)} & \colhead{(\kms)}}
\startdata
    50388.7556\dotfill  &   0.673 &    $+$108.89 &   $-$106.53 &   $+$1.19  &    $-$0.51   \nl 
    50403.8862\dotfill  &   0.254 &    $-$122.31 &   $+$119.75 &   $-$1.72  &    $-$1.26   \nl 
    50407.8678\dotfill  &   0.144 &    $-$100.42 &   $+$106.76 &   $-$5.72  &   $+$11.51   \nl 
    50409.9510\dotfill  &   0.702 &    $+$119.48 &   $-$109.13 &   $+$3.18  &    $+$5.45   \nl 
    50415.8348\dotfill  &   0.928 &     $+$51.76 &    $-$50.64 &   $-$1.76  &    $+$1.50   \nl 
    50420.8603\dotfill  &   0.100 &     $-$80.65 &    $+$65.08 &  $-$10.16  &    $-$6.10   \nl 
    50436.8784\dotfill  &   0.771 &    $+$122.97 &   $-$118.65 &   $+$2.31  &    $+$0.26   \nl 
    50438.8941\dotfill  &   0.247 &    $-$119.84 &   $+$120.52 &   $+$0.77  &    $-$0.50   \nl 
    50439.8135\dotfill  &   0.376 &     $-$81.36 &    $+$91.03 &   $+$3.44  &    $+$5.62   \nl 
    50440.8755\dotfill  &   0.680 &    $+$112.34 &   $-$108.93 &   $+$2.20  &    $-$0.48   \nl 
    50441.8847\dotfill  &   0.919 &     $+$57.85 &    $-$57.77 &   $-$1.58  &    $+$0.25   \nl 
    50442.9609\dotfill  &   0.241 &    $-$118.98 &   $+$120.83 &   $+$1.46  &    $-$0.02   \nl 
    50443.8319\dotfill  &   0.311 &    $-$114.27 &   $+$113.58 &   $-$2.30  &    $+$1.14   \nl 
    50443.8978\dotfill  &   0.391 &     $-$78.87 &    $+$75.03 &   $-$3.04  &    $-$1.47   \nl 
    50446.9365\dotfill  &   0.123 &     $-$87.88 &    $+$83.51 &   $-$3.70  &    $-$1.29   \nl 
    50448.8145\dotfill  &   0.430 &     $-$52.27 &    $+$59.24 &   $-$0.91  &    $+$7.08   \nl 
    50456.8140\dotfill  &   0.254 &    $-$120.98 &   $+$121.63 &   $-$0.37  &    $+$0.61   \nl 
    50460.8356\dotfill  &   0.192 &    $-$115.11 &   $+$109.70 &   $-$2.33  &    $-$3.53   \nl 
    50462.7592\dotfill  &   0.555 &     $+$47.03 &    $-$37.36 &   $+$5.66  &    $+$2.70   \nl 
    50467.7616\dotfill  &   0.698 &    $+$112.50 &   $-$112.22 &   $-$2.81  &    $+$1.37   \nl 
    50469.7291\dotfill  &   0.114 &     $-$80.16 &    $+$79.12 &   $-$0.99  &    $-$0.69   \nl 
    50472.7437\dotfill  &   0.816 &    $+$115.76 &   $-$110.48 &   $+$4.44  &    $-$0.86   \nl 
    50474.7172\dotfill  &   0.240 &    $-$123.13 &   $+$121.55 &   $-$2.73  &    $+$0.74   \nl 
    50481.7527\dotfill  &   0.880 &     $+$83.50 &    $-$81.33 &   $+$0.09  &    $+$0.54   \nl 
    50483.6753\dotfill  &   0.241 &    $-$119.27 &   $+$116.35 &   $+$1.18  &    $-$4.52   \nl 
    50485.6715\dotfill  &   0.693 &    $+$105.25 &   $-$113.22 &   $-$8.71  &    $-$0.97   \nl 
    50485.7188\dotfill  &   0.751 &    $+$128.69 &   $-$123.82 &   $+$6.97  &    $-$3.86   \nl 
    50486.5903\dotfill  &   0.821 &    $+$112.82 &   $-$107.42 &   $+$2.99  &    $+$0.72   \nl 
    50492.6584\dotfill  &   0.273 &    $-$113.04 &   $+$116.99 &   $+$6.31  &    $-$2.79   \nl 
    50493.4989\dotfill  &   0.305 &    $-$113.09 &   $+$108.08 &   $+$0.29  &    $-$5.75   \nl 
    50494.6927\dotfill  &   0.771 &    $+$119.26 &   $-$119.44 &   $-$1.36  &    $-$0.57   \nl 
    50495.6047\dotfill  &   0.891 &     $+$77.98 &    $-$73.99 &   $+$1.05  &    $+$1.44   \nl 
    50495.7929\dotfill  &   0.123 &     $-$87.16 &    $+$87.67 &   $-$3.34  &    $+$3.23   \nl 
    50495.7974\dotfill  &   0.128 &     $-$82.21 &    $+$85.03 &   $+$4.58  &    $-$2.36   \nl 
    50495.9059\dotfill  &   0.261 &    $-$119.41 &   $+$117.38 &   $+$0.92  &    $-$3.36   \nl 
    50496.6755\dotfill  &   0.206 &    $-$114.23 &   $+$119.85 &   $+$1.90  &    $+$3.28   \nl 
    50497.8476\dotfill  &   0.646 &     $+$98.46 &    $-$93.88 &   $+$1.75  &    $+$1.21   \nl 
    50499.8666\dotfill  &   0.125 &     $-$84.78 &    $+$84.18 &   $+$0.57  &    $-$1.78   \nl 
    50500.8001\dotfill  &   0.272 &    $-$117.25 &   $+$119.51 &   $+$2.25  &    $-$0.41   \nl 
    50502.7780\dotfill  &   0.701 &    $+$117.19 &   $-$114.98 &   $+$1.21  &    $-$0.73   \nl 
    50503.6378\dotfill  &   0.757 &    $+$121.12 &   $-$118.79 &   $-$0.49  &    $+$1.07   \nl 
    50503.6537\dotfill  &   0.776 &    $+$119.62 &   $-$120.45 &   $-$0.46  &    $-$2.12   \nl 
    50504.7441\dotfill  &   0.115 &     $-$75.68 &    $+$77.36 &   $+$4.10  &    $-$3.06   \nl 
    50507.7246\dotfill  &   0.776 &    $+$116.37 &   $-$115.63 &   $-$3.79  &    $+$2.78   \nl 
    50510.6167\dotfill  &   0.327 &    $-$106.24 &   $+$109.17 &   $+$0.38  &    $+$2.06   \nl 
    50514.6515\dotfill  &   0.282 &    $-$116.10 &   $+$118.30 &   $+$2.04  &    $-$0.27   \nl 
    50515.7804\dotfill  &   0.669 &    $+$102.70 &   $-$112.19 &   $-$3.56  &    $-$7.60   \nl 
    50521.5605\dotfill  &   0.767 &    $+$119.84 &   $-$124.51 &   $-$1.18  &    $-$5.24   \nl 
    50523.6368\dotfill  &   0.317 &    $-$106.03 &   $+$102.84 &   $+$4.03  &    $-$7.69   \nl 
    50525.5846\dotfill  &   0.709 &    $+$115.42 &   $-$107.33 &   $-$2.31  &    $+$8.66   \nl 
    50530.6164\dotfill  &   0.888 &     $+$83.37 &    $-$70.94 &   $+$4.68  &    $+$6.23   \nl 
    50531.7052\dotfill  &   0.226 &    $-$121.52 &   $+$122.04 &   $-$2.31  &    $+$2.41   \nl 
    50532.6553\dotfill  &   0.392 &     $-$72.17 &    $+$75.55 &   $+$3.13  &    $-$0.41   \nl 
    50534.5933\dotfill  &   0.772 &    $+$108.19 &   $-$119.05 &  $-$12.33  &    $-$0.28   \nl 
    50536.5972\dotfill  &   0.233 &    $-$123.94 &   $+$121.42 &   $-$3.97  &    $+$1.03   \nl 
    50538.5472\dotfill  &   0.628 &     $+$89.05 &    $-$83.23 &   $+$1.18  &    $+$3.07   \nl 
    50540.5477\dotfill  &   0.085 &     $-$62.16 &    $+$66.77 &   $-$1.12  &    $+$4.99   \nl 
    50541.6248\dotfill  &   0.408 &     $-$61.92 &    $+$70.87 &   $+$4.00  &    $+$4.23   \nl 
    50543.5521\dotfill  &   0.774 &    $+$116.19 &   $-$118.57 &   $-$4.10  &    $-$0.03   \nl 
    50550.5521\dotfill  &   0.371 &     $-$83.30 &    $+$92.27 &   $+$3.97  &    $+$4.40   \nl 
    50554.6000\dotfill  &   0.342 &     $-$97.76 &   $+$100.83 &   $+$3.14  &    $-$0.59   \nl 
    50565.5499\dotfill  &   0.789 &    $+$114.71 &   $-$112.88 &   $-$3.31  &    $+$3.41   \nl 
    50566.6741\dotfill  &   0.170 &    $-$107.29 &   $+$107.54 &   $-$1.63  &    $+$1.39   \nl 
    50566.6821\dotfill  &   0.180 &    $-$109.89 &   $+$110.58 &   $-$0.83  &    $+$1.05   \nl 
    50566.6898\dotfill  &   0.189 &    $-$114.38 &   $+$110.43 &   $-$2.45  &    $-$1.96   \nl 
    50569.5426\dotfill  &   0.693 &    $+$114.87 &   $-$113.97 &   $+$0.90  &    $-$1.71   \nl 
    50571.5514\dotfill  &   0.160 &     $-$99.28 &   $+$103.78 &   $+$2.38  &    $+$1.60   \nl 
    50580.5514\dotfill  &   0.212 &    $-$114.14 &   $+$116.26 &   $+$3.13  &    $-$1.44   \nl 
    50584.5512\dotfill  &   0.124 &     $-$87.01 &    $+$81.47 &   $-$2.16  &    $-$3.99   \nl 
    50589.5422\dotfill  &   0.254 &    $-$117.93 &   $+$123.74 &   $+$2.67  &    $+$2.72   \nl 
    50795.9108\dotfill  &   0.690 &    $+$112.24 &   $-$112.76 &   $-$0.97  &    $-$1.26   \nl 
    50796.8181\dotfill  &   0.804 &    $+$114.22 &   $-$116.23 &   $-$0.53  &    $-$3.19   \nl 
    50800.9569\dotfill  &   0.887 &     $+$82.74 &    $-$75.43 &   $+$3.20  &    $+$2.59   \nl 
    50802.8367\dotfill  &   0.196 &    $-$115.58 &   $+$108.42 &   $-$1.97  &    $-$5.64   \nl 
    50803.7832\dotfill  &   0.358 &     $-$94.08 &    $+$95.93 &   $-$0.24  &    $+$1.53   \nl 
    50804.9232\dotfill  &   0.758 &    $+$119.70 &   $-$120.27 &   $-$1.87  &    $-$0.45   \nl 
    50811.8377\dotfill  &   0.249 &    $-$120.25 &   $+$117.09 &   $+$0.38  &    $-$3.96   \nl 
    50816.8156\dotfill  &   0.363 &     $-$94.29 &    $+$93.13 &   $-$2.77  &    $+$1.04   \nl 
    50824.8106\dotfill  &   0.181 &    $-$116.98 &   $+$113.83 &   $-$7.51  &    $+$3.89   \nl 
    50827.7634\dotfill  &   0.807 &    $+$117.60 &   $-$118.11 &   $+$3.67  &    $-$5.90   \nl 
    50828.5942\dotfill  &   0.828 &    $+$107.84 &   $-$107.98 &   $+$0.27  &    $-$2.09   \nl 
    50828.9332\dotfill  &   0.244 &    $-$123.14 &   $+$122.29 &   $-$2.59  &    $+$1.33   \nl 
    50828.9376\dotfill  &   0.249 &    $-$122.88 &   $+$121.69 &   $-$2.25  &    $+$0.64   \nl 
    50828.9419\dotfill  &   0.255 &    $-$120.93 &   $+$122.29 &   $-$0.35  &    $+$1.29   \nl 
    50828.9462\dotfill  &   0.260 &    $-$122.45 &   $+$122.43 &   $-$2.05  &    $+$1.62   \nl 
    50835.8190\dotfill  &   0.700 &    $+$120.45 &   $-$108.32 &   $+$4.60  &    $+$5.81   \nl 
    50839.8816\dotfill  &   0.689 &    $+$110.27 &   $-$115.55 &   $-$2.79  &    $-$4.20   \nl 
    50840.7282\dotfill  &   0.729 &    $+$118.85 &   $-$120.53 &   $-$1.83  &    $-$1.60   \nl 
    50843.6844\dotfill  &   0.360 &     $-$90.45 &    $+$96.58 &   $+$2.58  &    $+$2.98   \nl 
    50845.6723\dotfill  &   0.801 &    $+$116.65 &   $-$112.75 &   $+$1.07  &    $+$1.11   \nl 
\enddata
\tablenotetext{a}{Referred to the time of eclipse given in the text (eq.[1]).}
\end{deluxetable}

\begin{deluxetable}{lc}
\tablenum{2}
\tablewidth{20pc}
\tablecaption{Spectroscopic orbital solution for YY~Gem.\label{tabspecelem}}
\tablehead{
\colhead{\hfil~~~~~~~~~~~~~~~Element~~~~~~~~~~~~~~~~} &  \colhead{Value} }
\startdata
\tablevspace{2pt}
\multispan{2}{Adjusted quantities\hfil} \nl
\tablevspace{2pt}
~~~$P$ (days)\tablenotemark{a}\dotfill               &  0.814282212                  \nl
~~~$\gamma$ (\kms)\dotfill                           &  $+0.54$~$\pm$~0.26\phm{$+$}  \nl
~~~$K_{\rm A}$ (\kms)\dotfill                        &  121.18~$\pm$~0.42\phn\phn    \nl
~~~$K_{\rm B}$ (\kms)\dotfill                        &  120.51~$\pm$~0.42\phn\phn    \nl
~~~$e$\tablenotemark{b}\dotfill                      &  0                            \nl
~~~Min~I (HJD$-$2,400,000)\tablenotemark{a}\dotfill  &  49,345.112327                \nl
\tablevspace{2pt}
\multispan{2}{Derived quantities\hfil} \nl
\tablevspace{2pt}
~~~$M_{\rm A}\sin^3 i$ (M$_{\sun}$)\dotfill          &  0.5938~$\pm$~0.0046          \nl
~~~$M_{\rm B}\sin^3 i$ (M$_{\sun}$)\dotfill          &  0.5971~$\pm$~0.0046          \nl
~~~$q\equiv M_{\rm B}/M_{\rm A}$\dotfill             &  1.0056~$\pm$~0.0050          \nl
~~~$a_{\rm A}\sin i$ (10$^6$ km)\dotfill             &  1.3569~$\pm$~0.0047          \nl
~~~$a_{\rm B}\sin i$ (10$^6$ km)\dotfill             &  1.3493~$\pm$~0.0047          \nl
~~~$a \sin i$ (R$_{\sun}$)\dotfill                   &  3.8882~$\pm$~0.0095          \nl
\tablevspace{2pt}
\multispan{2}{Other quantities pertaining to the fit\hfil} \nl
\tablevspace{2pt}
~~~$N_{\rm obs}$\dotfill                             & 90                            \nl
~~~Time span (days)\dotfill                          & 456.9                         \nl
~~~$\sigma_{\rm A}$ (\kms)\dotfill                   & 3.44                          \nl
~~~$\sigma_{\rm B}$ (\kms)\dotfill                   & 3.44                          \nl
\enddata
\tablenotetext{a}{Adopted from the ephemeris in \S3, and held fixed.}
\tablenotetext{b}{Circular orbit adopted (see text).}
\end{deluxetable}

\begin{deluxetable}{ccccrcc}
\tablewidth{35pc}
\tablenum{3}
\tablecaption{Eclipse timings and residuals for YY~Gem.\label{tabtmin}}
\tablehead{
\colhead{HJD} & \colhead{}  & \colhead{}  &  \colhead{} & \colhead{} &
\colhead{(O-C)}  & \colhead{} \\
\colhead{\hbox{~~(2,400,000+)~~}} & \colhead{Year}  &  \colhead{Eclipse\tablenotemark{a}} &
\colhead{Type\tablenotemark{b}} & \colhead{Epoch\tablenotemark{c}} & 
\colhead{(days)} & \colhead{Reference}
}
\startdata
 24500.5471\dotfill  &  1925.9563  &  1  &  pg  &  $-$30511.0  &  $-$0.0007  &   \phn1 \nl
 24573.4233\dotfill  &  1926.1558  &  2  &  pg  &  $-$30421.5  &  $-$0.0027  &   \phn1 \nl
 24584.4125\dotfill  &  1926.1859  &  1  &  pg  &  $-$30408.0  &  $-$0.0063  &   \phn1 \nl
 24591.3347\dotfill  &  1926.2049  &  2  &  pg  &  $-$30399.5  &  $-$0.0055  &   \phn1 \nl
 24595.4127\tablenotemark{d}\dotfill  &  1926.2161  &  2  &  pg  &  $-$30394.5  &  $+$0.0011  &   \phn1 \nl
 24619.4308\dotfill  &  1926.2818  &  1  &  pg  &  $-$30365.0  &  $-$0.0022  &   \phn1 \nl
 24639.3854\dotfill  &  1926.3364  &  2  &  pg  &  $-$30340.5  &  $+$0.0025  &   \phn1 \nl
 24791.6548\dotfill  &  1926.7533  &  2  &  pg  &  $-$30153.5  &  $+$0.0011  &   \phn2 \nl
 24848.6537\dotfill  &  1926.9094  &  2  &  pg  &  $-$30083.5  &  $+$0.0003  &   \phn2 \nl
 24875.5268\dotfill  &  1926.9830  &  2  &  pg  &  $-$30050.5  &  $+$0.0021  &   \phn2 \nl
 24916.6466\dotfill  &  1927.0955  &  1  &  pg  &  $-$30000.0  &  $+$0.0006  &   \phn2 \nl
 24920.3112\dotfill  &  1927.1056  &  2  &  pg  &  $-$29995.5  &  $+$0.0010  &   \phn2 \nl
 24921.5306\dotfill  &  1927.1089  &  1  &  pg  &  $-$29994.0  &  $-$0.0011  &   \phn2 \nl
 24922.3441\dotfill  &  1927.1111  &  1  &  pg  &  $-$29993.0  &  $-$0.0018  &   \phn2 \nl
 24961.4304\dotfill  &  1927.2182  &  1  &  pg  &  $-$29945.0  &  $-$0.0011  &   \phn2 \nl
 25230.5519\dotfill  &  1927.9550  &  2  &  pg  &  $-$29614.5  &  $+$0.0001  &   \phn2 \nl
 25234.6211\dotfill  &  1927.9661  &  2  &  pg  &  $-$29609.5  &  $-$0.0021  &   \phn2 \nl
 25242.3568\dotfill  &  1927.9873  &  1  &  pg  &  $-$29600.0  &  $-$0.0021  &   \phn2 \nl
 25687.3656\dotfill  &  1929.2057  &  2  &  pg  &  $-$29053.5  &  $+$0.0015  &   \phn2 \nl
 25698.3561\dotfill  &  1929.2357  &  1  &  pg  &  $-$29040.0  &  $-$0.0008  &   \phn2 \nl
 27158.3641\dotfill  &  1933.2330  &  1  &  pg  &  $-$27247.0  &  $-$0.0008  &   \phn3 \nl
 27160.4011\dotfill  &  1933.2386  &  2  &  pg  &  $-$27244.5  &  $+$0.0005  &   \phn3 \nl
 27461.2782\dotfill  &  1934.0624  &  1  &  pg  &  $-$26875.0  &  $+$0.0003  &   \phn4 \nl
 28545.4929\dotfill  &  1937.0308  &  2  &  pg  &  $-$25543.5  &  $-$0.0017  &   \phn3 \nl
 28571.5540\dotfill  &  1937.1021  &  2  &  pg  &  $-$25511.5  &  $+$0.0023  &   \phn3 \nl
 28596.3861\dotfill  &  1937.1701  &  1  &  pg  &  $-$25481.0  &  $-$0.0012  &   \phn3 \nl
 29639.4827\dotfill  &  1940.0260  &  1  &  pg  &  $-$24200.0  &  $-$0.0001  &   \phn3 \nl
 30466.3861\dotfill  &  1942.2899  &  2  &  pg  &  $-$23184.5  &  $-$0.0003  &   \phn3 \nl
 32605.9146\dotfill  &  1948.1476  &  1  &  pe  &  $-$20557.0  &  $+$0.0017  &   \phn5 \nl
 32606.3222\dotfill  &  1948.1487  &  2  &  pe  &  $-$20556.5  &  $+$0.0022  &   \phn5 \nl
 32965.8275\dotfill  &  1949.1330  &  1  &  pe  &  $-$20115.0  &  $+$0.0019  &   \phn5 \nl
 32966.2353\dotfill  &  1949.1341  &  2  &  pe  &  $-$20114.5  &  $+$0.0025  &   \phn5 \nl
 40252.4400\dotfill  &  1969.0827  &  2  &  v   &  $-$11166.5  &  $+$0.0100  &   \phn6 \nl
 40259.3550\dotfill  &  1969.1016  &  1  &  v   &  $-$11158.0  &  $+$0.0036  &   \phn6 \nl
 40274.4220\dotfill  &  1969.1428  &  2  &  v   &  $-$11139.5  &  $+$0.0064  &   \phn6 \nl
 40283.3800\dotfill  &  1969.1674  &  2  &  v   &  $-$11128.5  &  $+$0.0073  &   \phn6 \nl
 40316.3570\dotfill  &  1969.2577  &  1  &  v   &  $-$11088.0  &  $+$0.0058  &   \phn6 \nl
 40316.3620\dotfill  &  1969.2577  &  1  &  v   &  $-$11088.0  &  $+$0.0108  &   \phn6 \nl
 40322.4170\dotfill  &  1969.2742  &  2  &  v   &  $-$11080.5  & ($-$0.0413) &   \phn6 \nl
 40353.3860\dotfill  &  1969.3590  &  2  &  v   &  $-$11042.5  &  $-$0.0150  &   \phn6 \nl
 40353.4100\dotfill  &  1969.3591  &  2  &  v   &  $-$11042.5  &  $+$0.0090  &   \phn6 \nl
 40561.4590\dotfill  &  1969.9287  &  1  &  v   &  $-$10787.0  &  $+$0.0089  &   \phn6 \nl
 40589.5620\dotfill  &  1970.0056  &  2  &  v   &  $-$10752.5  &  $+$0.0192  &   \phn6 \nl
 40658.3540\dotfill  &  1970.1940  &  1  &  v   &  $-$10668.0  &  $+$0.0043  &   \phn6 \nl
 40854.5810\dotfill  &  1970.7312  &  1  &  v   &  $-$10427.0  &  $-$0.0107  &   \phn6 \nl
 40938.4700\dotfill  &  1970.9609  &  1  &  v   &  $-$10324.0  &  $+$0.0072  &   \phn6 \nl
 40968.9953\dotfill  &  1971.0445  &  2  &  pe  &  $-$10286.5  & ($-$0.0031) &   \phn7 \nl
 40969.8116\dotfill  &  1971.0467  &  2  &  pe  &  $-$10285.5  & ($-$0.0010) &   \phn7 \nl
 40970.6265\dotfill  &  1971.0489  &  2  &  pe  &  $-$10284.5  & ($-$0.0004) &   \phn7 \nl
 40971.8466\dotfill  &  1971.0523  &  1  &  pe  &  $-$10283.0  & ($-$0.0017) &   \phn7 \nl
 41024.3810\dotfill  &  1971.1961  &  2  &  v   &  $-$10218.5  &  $+$0.0115  &   \phn6 \nl
 41315.8847\dotfill  &  1971.9942  &  2  &  v   &   $-$9860.5  &  $+$0.0021  &   \phn8 \nl
 41316.2979\dotfill  &  1971.9953  &  1  &  v   &   $-$9860.0  &  $+$0.0082  &   \phn8 \nl
 41401.3920\dotfill  &  1972.2283  &  2  &  v   &   $-$9755.5  &  $+$0.0098  &   \phn9 \nl
 41410.3400\dotfill  &  1972.2528  &  2  &  v   &   $-$9744.5  &  $+$0.0007  &   \phn9 \nl
 41416.4580\dotfill  &  1972.2696  &  1  &  v   &   $-$9737.0  &  $+$0.0116  &   \phn9 \nl
 41434.3660\dotfill  &  1972.3186  &  1  &  v   &   $-$9715.0  &  $+$0.0054  &  10 \nl
 41436.3880\dotfill  &  1972.3241  &  2  &  v   &   $-$9712.5  &  $-$0.0083  &   \phn9 \nl
 41681.0835\dotfill  &  1972.9941  &  1  &  v   &   $-$9412.0  &  $-$0.0047  &   \phn8 \nl
 41719.3500\dotfill  &  1973.0988  &  1  &  v   &   $-$9365.0  &  $-$0.0094  &  11 \nl
 41789.3880\dotfill  &  1973.2906  &  1  &  v   &   $-$9279.0  &  $+$0.0003  &  12 \nl
 42035.3260\dotfill  &  1973.9639  &  1  &  v   &   $-$8977.0  &  $+$0.0251  &  13 \nl
 42064.2100\dotfill  &  1974.0430  &  2  &  pe  &   $-$8941.5  &  $+$0.0021  &  14 \nl
 42403.3570\dotfill  &  1974.9715  &  1  &  v   &   $-$8525.0  &  $+$0.0005  &  15 \nl
 42411.4956\dotfill  &  1974.9938  &  1  &  v   &   $-$8515.0  &  $-$0.0037  &   \phn8 \nl
 42411.9100\dotfill  &  1974.9950  &  2  &  v   &   $-$8514.5  &  $+$0.0036  &   \phn8 \nl
 42416.3990\dotfill  &  1975.0073  &  1  &  v   &   $-$8509.0  &  $+$0.0140  &  16 \nl
 42430.2300\dotfill  &  1975.0451  &  1  &  pe  &   $-$8492.0  &  $+$0.0022  &  14 \nl
 42453.0300\dotfill  &  1975.1075  &  1  &  pe  &   $-$8464.0  &  $+$0.0023  &  14 \nl
 42464.4290\dotfill  &  1975.1388  &  1  &  v   &   $-$8450.0  &  $+$0.0014  &  17 \nl
 42467.2810\dotfill  &  1975.1466  &  2  &  v   &   $-$8446.5  &  $+$0.0034  &  17 \nl
 42469.3000\dotfill  &  1975.1521  &  1  &  v   &   $-$8444.0  &  $-$0.0133  &  17 \nl
 42829.6340\dotfill  &  1976.1386  &  2  &  pe  &   $-$8001.5  &  $+$0.0008  &  18 \nl
 42835.3340\dotfill  &  1976.1542  &  2  &  v   &   $-$7994.5  &  $+$0.0008  &  19 \nl
 42837.3760\dotfill  &  1976.1598  &  1  &  v   &   $-$7992.0  &  $+$0.0071  &  19 \nl
 43131.3280\dotfill  &  1976.9646  &  1  &  v   &   $-$7631.0  &  $+$0.0032  &  20 \nl
 43142.7276\dotfill  &  1976.9958  &  1  &  v   &   $-$7617.0  &  $+$0.0029  &   \phn8 \nl
 43510.3630\dotfill  &  1978.0024  &  2  &  v   &   $-$7165.5  &  $-$0.0101  &  21 \nl
 43514.8660\dotfill  &  1978.0147  &  1  &  pe  &   $-$7160.0  &  $+$0.0143  &  22 \nl
 43949.6790\dotfill  &  1979.2051  &  1  &  pe  &   $-$6626.0  &  $+$0.0006  &  23 \nl
 43960.6731\dotfill  &  1979.2352  &  2  &  pe  &   $-$6612.5  &  $+$0.0019  &  23 \nl
 43969.6277\dotfill  &  1979.2598  &  2  &  pe  &   $-$6601.5  &  $-$0.0006  &  23 \nl
 44295.3440\dotfill  &  1980.1515  &  2  &  v   &   $-$6201.5  &  $+$0.0028  &  24 \nl
 44303.4950\dotfill  &  1980.1738  &  2  &  v   &   $-$6191.5  &  $+$0.0110  &  25 \nl
 44343.3880\dotfill  &  1980.2831  &  2  &  v   &   $-$6142.5  &  $+$0.0042  &  25 \nl
 44637.3410\dotfill  &  1981.0879  &  2  &  pe  &   $-$5781.5  &  $+$0.0013  &  26 \nl
 44709.4065\dotfill  &  1981.2852  &  1  &  pe  &   $-$5693.0  &  $+$0.0028  &  27 \nl
 45036.3430\dotfill  &  1982.1803  &  2  &  v   &   $-$5291.5  &  $+$0.0050  &  28 \nl
 45036.3500\dotfill  &  1982.1803  &  2  &  v   &   $-$5291.5  &  $+$0.0120  &  28 \nl
 46017.5488\dotfill  &  1984.8667  &  2  &  pe  &   $-$4086.5  &  $+$0.0007  &  29 \nl
 46018.7683\dotfill  &  1984.8700  &  1  &  pe  &   $-$4085.0  &  $-$0.0012  &  30 \nl
 46019.1766\dotfill  &  1984.8711  &  2  &  pe  &   $-$4084.5  &  $-$0.0000  &  30 \nl
 46019.5800\dotfill  &  1984.8722  &  1  &  pe  &   $-$4084.0  &  $-$0.0038  &  31 \nl
 46121.3770\dotfill  &  1985.1509  &  1  &  v   &   $-$3959.0  &  $+$0.0079  &  32 \nl
 46474.3550\dotfill  &  1986.1173  &  2  &  v   &   $-$3525.5  &  $-$0.0054  &  33 \nl
 46533.3960\dotfill  &  1986.2790  &  1  &  v   &   $-$3453.0  &  $+$0.0001  &  33 \nl
 46825.3190\dotfill  &  1987.0782  &  2  &  v   &   $-$3094.5  &  $+$0.0030  &  34 \nl
 46827.3550\dotfill  &  1987.0838  &  1  &  v   &   $-$3092.0  &  $+$0.0033  &  34 \nl
 46860.3330\dotfill  &  1987.1741  &  2  &  v   &   $-$3051.5  &  $+$0.0028  &  34 \nl
 47182.3800\dotfill  &  1988.0558  &  1  &  v   &   $-$2656.0  &  $+$0.0012  &   \phn6 \nl
 47208.4300\dotfill  &  1988.1271  &  1  &  v   &   $-$2624.0  &  $-$0.0058  &  35 \nl
 47592.3680\dotfill  &  1989.1783  &  2  &  v   &   $-$2152.5  &  $-$0.0019  &  36 \nl
 48344.3550\dotfill  &  1991.2371  &  1  &  v   &   $-$1229.0  &  $-$0.0045  &  37 \nl
 48357.3760\dotfill  &  1991.2728  &  1  &  v   &   $-$1213.0  &  $-$0.0120  &  37 \nl
 48618.3630\dotfill  &  1991.9873  &  2  &  v   &    $-$892.5  &  $-$0.0025  &  38 \nl
 48686.3530\dotfill  &  1992.1735  &  1  &  v   &    $-$809.0  &  $-$0.0050  &  38 \nl
 48743.3530\dotfill  &  1992.3295  &  1  &  v   &    $-$739.0  &  $-$0.0048  &  38 \nl
 50113.3820\dotfill  &  1996.0804  &  2  &  v   &     943.5  &  $-$0.0056  &  39 \nl
 50146.3680\dotfill  &  1996.1708  &  1  &  v   &     984.0  &  $+$0.0020  &  39 \nl
 50170.3830\dotfill  &  1996.2365  &  2  &  v   &    1013.5  &  $-$0.0044  &  39 \nl
 50192.3710\dotfill  &  1996.2967  &  2  &  v   &    1040.5  &  $-$0.0020  &  39 \nl
 50896.3199\dotfill  &  1998.2240  &  1  &  pe  &    1905.0  &  $-$0.0000  &  40 \nl
 51163.4047\dotfill  &  1998.9552  &  1  &  pe  &    2233.0  &  $+$0.0002  &  40 \nl
 51165.4400\dotfill  &  1998.9608  &  2  &  pe  &    2235.5  &  $-$0.0002  &  40 \nl
 51254.6030\dotfill  &  1999.2049  &  1  &  pe  &    2345.0  &  $-$0.0011  &  41 \nl
 51288.3920\dotfill  &  1999.2974  &  2  &  v   &    2386.5  &  $-$0.0048  &  42 \nl
\enddata
\tablenotetext{a}{1 = primary eclipse; 2 = secondary eclipse.}
\tablenotetext{b}{pg $=$ photographic, v $=$ visual, pe $=$ photoelectric.}
\tablenotetext{c}{In cycles, counted from epoch of primary minimum given in the text.}
\tablenotetext{d}{This time of minimum has often been mistakenly
listed by other investigators as 2,424,595.4105, which is actually a
\emph{mean} epoch derived by van Gent (1926) from his own period
analysis. The correct value to use is the actual observed epoch, as
listed here.}
\tablerefs{
(1) van Gent 1926; 
(2) van Gent 1931; 
(3) Binnendijk 1950;
(4) Gadomski 1934;
(5) Kron 1952;
(6) Kundera 2001;
(7) Leung \& Schneider 1978;
(8) Mallama 1980a;
(9) Peter 1972;
(10) Diethelm 1972;
(11) Diethelm 1973a;
(12) Diethelm 1973b;
(13) Diethelm 1974;
(14) Budding 1975 (see also Haisch et al.\ 1980);
(15) Diethelm 1975a;
(16) Diethelm 1975b;
(17) Diethelm 1975c;
(18) Mallama et al.\ 1977;
(19) Braune \& H\"ubscher 1979;
(20) Diethelm 1977;
(21) Diethelm 1978;
(22) Kodaira \& Ichimura 1980;
(23) Mallama 1980b;
(24) Diethelm 1980a;
(25) Diethelm 1980b;
(26) Diethelm 1981;
(27) Braune \& Mundry 1981;
(28) Braune \& Mundry 1982;
(29) Geyer \& Kamper 1985;
(30) Haisch et al.\ 1990;
(31) Zsoldos 1986;
(32) Diethelm 1985;
(33) H\"ubscher, Lichtenknecker \& Meyer 1987;
(34) Braune \& H\"ubscher 1987;
(35) Diethelm 1988;
(36) H\"ubscher, Lichtenknecker \& Wunder 1990;
(37) H\"ubscher, Agerer \& Wunder 1991;
(38) H\"ubscher, Agerer \& Wunder 1992;
(39) H\"ubscher \& Agerer 1996;
(40) Agerer \& H\"ubscher 1999;
(41) Sowell et al.\ 2001;
(42) H\"ubscher et al.\ 1999.}
\end{deluxetable}

\begin{deluxetable}{lcc}
\tablenum{4}
\tablewidth{0pc}
\small
\tablecaption{Results from the light curve solutions for YY~Gem.\label{tablightelem}}
\tablehead{
\colhead{} & \colhead{LS78 data} & \colhead{Kron (1952) data} \\
\colhead{~~~~~~~~~~~~~~~~Parameter~~~~~~~~~~~~~~~~~} & \colhead{($V\!RI$)} & \colhead{($V$)}
}
\startdata
\multicolumn{3}{l}{Geometric and radiative parameters}  \nl
~~~$P$ (days) (fixed)\tablenotemark{a}\dotfill             & 0.814282212   & 0.814282212 \nl
~~~$\Delta\phi$\dotfill                   & $-0.0016\pm0.0002$\phm{$-$}   &  $+0.0025\pm0.0010$\phm{$+$} \nl
~~~$e$ (fixed)\dotfill                    &  0            &  0          \nl
~~~$q\equiv M_B/M_A$ (fixed)\tablenotemark{b}\dotfill            & 1.0056       & 1.0056     \nl
~~~$\Omega_A$ (fixed)\dotfill             & 7.315         & 7.48        \nl
~~~$\Omega_B$\dotfill                     & 7.334 $\pm$ 0.042  &  7.50 $\pm$ 0.40 \nl
~~~$r_{\rm point}$\dotfill  &  0.1600  &  0.1558 \nl
~~~$r_{\rm pole}$\dotfill  &  0.1583  &  0.1543 \nl
~~~$r_{\rm side}$\dotfill  &  0.1589  &  0.1548 \nl
~~~$r_{\rm back}$\dotfill  &  0.1598  &  0.1556 \nl
~~~$r = r_A = r_B$\tablenotemark{c}\dotfill  & 0.1590 $\pm$ 0.0014  &  0.1549 $\pm$ 0.0090  \nl
~~~$i$ (deg)\dotfill                      & 86.29 $\pm$ 0.10\phn     &  86.39 $\pm$ 0.54\phn  \nl
~~~$T_{\rm eff}^A$ (K) (fixed)\dotfill & 3820       & 3820        \nl
~~~$T_{\rm eff}^B/T_{\rm eff}^A$\dotfill    & 0.984 $\pm$ 0.004  &  1.003 $\pm$ 0.012 \nl
~~~$L_B/L_A$ ($V$ band)\dotfill           & 0.886 $\pm$ 0.010    &  1.021 $\pm$ 0.074 \nl
~~~$L_B/L_A$ ($R$ band)\dotfill           & 0.894 $\pm$ 0.010    &  \nodata           \nl
~~~$L_B/L_A$ ($I$ band)\dotfill           & 0.918 $\pm$ 0.010    &  \nodata           \nl
~~~$L_3$ (fixed)\dotfill                  &  0                   &  0                 \nl
~~~Albedo (fixed)\dotfill                 & 0.5           & 0.5         \nl
~~~Gravity brightening (fixed)\dotfill      & 0.2           & 0.2         \nl
\tablevspace{3pt}
\multicolumn{3}{l}{Limb darkening coefficients (Logarithmic law)}  \nl
~~~$x_A$ and $y_A$ ($V$ band)\dotfill     & 0.814~,~0.303        &  0.814~,~0.303     \nl
~~~$x_B$ and $y_B$ ($V$ band)\dotfill     & 0.815~,~0.331        &  0.813~,~0.306     \nl
~~~$x_A$ and $y_A$ ($R$ band)\dotfill     & 0.777~,~0.312        &  \nodata           \nl
~~~$x_B$ and $y_B$ ($R$ band)\dotfill     & 0.776~,~0.327        &  \nodata           \nl
~~~$x_A$ and $y_A$ ($I$ band)\dotfill     & 0.689~,~0.354        &  \nodata           \nl
~~~$x_B$ and $y_B$ ($I$ band)\dotfill     & 0.689~,~0.368        &  \nodata           \nl
\tablevspace{3pt}
\multicolumn{1}{l}{Spot parameters}       &                      &                    \nl
~~~$T_{\rm spot}/T_{\rm phot}$ (fixed)\dotfill &  0.90  & 0.90        \nl
~~~Latitude (deg) (fixed)\dotfill         &  $+$45          &   $+$45       \nl
\tablevspace{3pt}
~~~Phase for spot \#1\dotfill             &  0.733               & 0.516              \nl
~~~Radius for spot \#1 (deg)\dotfill      &  20.1                & 25.5               \nl
~~~Location of spot \#1\dotfill           &  Primary             & Secondary             \nl
\tablevspace{3pt}
~~~Phase for spot \#2\dotfill             &  0.587               & 0.836              \nl
~~~Radius for spot \#2 (deg)\dotfill      &  21.0                & 35.0               \nl
~~~Location of spot \#2\dotfill           &  Secondary           & Secondary          \nl
\tablevspace{3pt}
~~~Phase for spot \#3\dotfill             &  0.250               & \nodata            \nl
~~~Radius for spot \#3 (deg)\dotfill      &  9.0                 & \nodata            \nl
~~~Location of spot \#3\dotfill           &  Primary             & \nodata            \nl
\tablevspace{3pt}
\multicolumn{1}{l}{Residuals from the fits}     &                      &                    \nl
~~~$\sigma_V$ (mag)\dotfill               & 0.013                & 0.020              \nl
~~~$\sigma_R$ (mag)\dotfill               & 0.011                & \nodata            \nl
~~~$\sigma_I$ (mag)\dotfill               & 0.009                & \nodata            \nl
\enddata
\tablenotetext{a}{Period from ephemeris in eq.(1).}
\tablenotetext{b}{Mass ratio from spectroscopic solution (see Table~2).}
\tablenotetext{c}{Fractional radius of a sphere with the same volume as the Roche equipotential.}
\end{deluxetable}
 
\begin{deluxetable}{lc}
\tablenum{5}
\tablewidth{19pc}
\tablecaption{Physical properties of the mean component of YY~Gem.\label{tabphys}}
\tablehead{
\colhead{~~~~~~~~~~Parameter~~~~~~~~~~} & \colhead{Value}}
\startdata
Mass (M$_{\sun}$)\dotfill            &  0.5992 $\pm$ 0.0047  \nl
Radius (R$_{\sun}$)\dotfill          &  0.6191 $\pm$ 0.0057  \nl
$\log g$ (cgs)\dotfill                     &  4.6317 $\pm$ 0.0083    \nl
$\bar\rho$ (gr~cm$^{-3}$)\dotfill    &  3.56 $\pm$ 0.10    \nl
$v \sin i$ (km~s$^{-1}$)\tablenotemark{a}\dotfill      &  37 $\pm$ 2\phn       \nl
$v_{\rm sync} \sin i$ (km~s$^{-1}$)\tablenotemark{b}\dotfill  &  38.5 $\pm$ 0.4\phn   \nl
$T_{\rm eff}$ (K)\dotfill            &  3820 $\pm$ 100\phn    \nl
$L/L_{\sun}$\tablenotemark{c}\dotfill            &  0.0733 $\pm$ 0.0015  \nl
$M_{\rm bol}$ (mag)\tablenotemark{c,d}\dotfill          &  7.569 $\pm$ 0.020    \nl
$M_V$ (mag)\tablenotemark{e}\dotfill                  &  8.950 $\pm$ 0.029    \nl
\enddata
\tablenotetext{a}{Measured projected rotational velocity.}
\tablenotetext{b}{Projected rotational velocity expected for synchronous rotation.}
\tablenotetext{c}{Computed from the radius and effective temperature.}
\tablenotetext{d}{Assumes $M_{\rm bol}^{\sun} = 4.73$.}
\tablenotetext{e}{Computed from the apparent visual magnitude of the system and the Hipparcos parallax of Castor, revised as described in the Appendix.}
\end{deluxetable}

\begin{deluxetable}{lcccl}
\tablewidth{35pc}
\tablenum{6}
\tablecaption{Effective temperature estimates for the mean component
of YY~Gem.\label{tabteff}}
\tablehead{
\colhead{Method\tablenotemark{a}} & \colhead{Color index} & \colhead{Sources\tablenotemark{b}} &
\colhead{Result (K)} & \colhead{Calibration used}}
\startdata
\bv        &  1.454  & 2,3,4,5,7  &  3877   & Arribas \& Mart\'\i nez Roger 1989 \nl
(\ri)$_C$  &  0.993  & 1,2,4,7,8  &  3723   & Bessell 1979 \nl
(\vi)$_C$  &  1.919  & 2,4,7,8    &  3719   & Bessell 1979 \nl
\vk        &  3.890  & 2,4        &  3841   & Arribas \& Mart\'\i nez Roger 1989 \nl
\by        &  0.794  & 6          &  3971   & Olsen 1984 \nl
SED        & \nodata & \nodata    &  3800   & Bopp et al.\ 1974 \nl
\enddata
\tablenotetext{a}{See text for details.}
\tablenotetext{b}{All individual values from the sources listed have been converted
to a uniform system following Leggett (1992).}
\tablerefs{
(1) Kron, Gascoigne \& White 1957;
(2) Johnson 1965;
(3) Eggen 1968;
(4) Veeder 1974;
(5) Budding 1975;
(6) Hilditch \& Hill 1975;
(7) Barnes, Evans \& Moffett 1978;
(8) Leung \& Schneider 1978
.}
\end{deluxetable}

\begin{deluxetable}{lcc}
\tablenum{7}
\tablewidth{23pc}
\tablecaption{Physical properties of the Hyades eclipsing binary V818~Tau.\label{tabvb22}}
\tablehead{
\colhead{~~~~~~~Parameter\tablenotemark{a}~~~~~~~} & \colhead{Primary} & \colhead{Secondary}
}
\startdata
Mass (M$_{\sun}$)\dotfill      &  1.0591~$\pm$~0.0062 & 0.7605~$\pm$~0.0062   \nl
Radius (R$_{\sun}$)\dotfill    &  0.900~$\pm$~0.016   & 0.768~$\pm$~0.010   \nl
$\log g$\dotfill               &  4.554~$\pm$~0.016   & 4.548~$\pm$~0.011   \nl
$T_{\rm eff}$ (K)\tablenotemark{b}\dotfill      &  5530~$\pm$~100\phn  & 4220~$\pm$~150\phn   \nl
$L/L_{\sun}$\dotfill           &  0.678~$\pm$~0.055   & 0.168~$\pm$~0.024   \nl
$M_V$ (mag)\tablenotemark{c}\dotfill            &  5.10~$\pm$~0.13     & 7.39~$\pm$~0.14   \nl
\enddata
\tablenotetext{a}{Based on the light-curve solution by Schiller \&
Milone (1987) and the spectroscopic analysis by Peterson \& Solensky
(1988).}
\tablenotetext{b}{Derived from photometric indices, as described in the text.}
\tablenotetext{c}{The parallax adopted is from the Hipparcos mission.}
\end{deluxetable}

\begin{deluxetable}{lcccccc@{}cc@{}c}
\tablewidth{42pc}
\tablenum{8}
\tablecaption{Age and metallicity estimates for Castor AB from
isochrone fitting.\label{tabagemet}}
\tablehead{
\colhead{} & \colhead{} & \colhead{Age} & \colhead{} & \colhead{} & \colhead{Mixing Length/} &
\colhead{$M_{\rm Aa}$} & \colhead{$M_{\rm Ba}$} & \colhead{$R_{\rm Aa}$} &
\colhead{$R_{\rm Ba}$} \\
\colhead{Model~/~BC~~~~~~~~~~~} & \colhead{Refs.} & \colhead{(Myr)} & \colhead{$Z$} &
\colhead{$Y$} & \colhead{Overshooting} & \colhead{(M$_{\sun}$)} &
\colhead{(M$_{\sun}$)} & \colhead{(R$_{\sun}$)} & \colhead{(R$_{\sun}$)}
}

\startdata
Girardi/Bertelli\dotfill  &  1 / 2 &  333 &  0.0198 &  0.275 &  1.68 / 0.25 &  2.30 &  1.80 &  2.24 & 1.71  \nl
\tablevspace{5pt}
Yi/Lejeune\dotfill     &  3 / 4 &  378 &  0.0175 &  0.265 &  1.74 / 0.20 &  2.29 &  1.80 &  2.20 & 1.68  \nl
Yi/Green\dotfill       &  3 / 5 &  369 &  0.0183 &  0.267 &  1.74 / 0.20 &  2.31 &  1.82 &  2.22 & 1.69  \nl
\tablevspace{5pt}
Claret/Flower\dotfill     &  6 / 7 &  410 &  0.0165 &  0.270 &  1.52 / 0.20 &  2.26 &  1.80 &  2.19 & 1.69  \nl
Claret/Flower\dotfill     &  6 / 7 &  381 &  0.0183 &  0.300 &  1.52 / 0.20 &  2.20 &  1.75 &  2.19 & 1.69  \nl
\tablevspace{5pt}
Schaller/Flower\dotfill   &  8 / 7 &  367 &  0.0182 &  0.295 &  1.60 / 0.20 &  2.26 &  1.79 &  2.18 & 1.68  \nl
\tablevspace{5pt}
Siess/Siess\dotfill       &  9 / 9 &  300 &  0.0189 &  0.275 &  1.50 / 0.00 &  2.34 &  1.85 &  2.15 & 1.67  \nl
Siess/Kenyon\tablenotemark{a}\dotfill    & \phn9 / 10 & 315 &  0.0186 &  0.274 &  1.50 / 0.00 &  2.36 &  1.84 &  2.20 & 1.67  \nl
Siess/Flower\dotfill      &  9 / 7 &  339 &  0.0174 &  0.272 &  1.50 / 0.00 &  2.29 &  1.82 &  2.11 & 1.64  \nl
\enddata
\tablenotetext{a}{Zero point of bolometric corrections adjusted for consistency with $M^{\sun}_{\rm bol}$.}
\tablerefs{
(1) Girardi et al.\ 2000;
(2) Bertelli et al.\ 1994;
(3) Yi et al.\ 2001;
(4) Lejeune et al.\ 1998;
(5) Green, Demarque \& King 1987;
(6) Claret \& Gim\'enez 1998;
(7) Flower 1996;
(8) Schaller et al.\ 1992;
(9) Siess, Forestini \& Dougados 1997;
(10) Kenyon \& Hartmann 1995.
}
\end{deluxetable}

\begin{deluxetable}{lcccc}
\tablewidth{32pc}
\tablenum{9}
\tablecaption{Summary of orbital and physical properties of Castor as a
sextuple stellar system.\label{tabtriple}}
\tablehead{\colhead{~~~~~~~~~~~~Parameter~~~~~~~~~~~~~} & \colhead{A} & 
\colhead{B} & \colhead{C} & \colhead{AB}}
\startdata
Mass (M$_{\sun}$)\tablenotemark{a}\dotfill & 2.27 $+$ 0.45 &
1.79 $+$ 0.32 & 0.60 $+$ 0.60 & 4.83 \\
Orbital inclination (deg)\tablenotemark{a,b}\dotfill & $\sim$28 & $\sim$90: & 86.3 & 114.5 \\
Orbital period\dotfill & 9.21280 d & 2.928315 d & 0.814282212 d & 467 yr \\  
Orbital eccentricity\dotfill & 0.4992 & 0 & 0 & 0.343\\
\enddata
\tablenotetext{a}{Model-dependent for Castor A and B.}
\tablenotetext{b}{Alternate values of the inclination angle for the orbits of Castor A, B, and C given
by $180\arcdeg - i$ are also possible.}
\end{deluxetable}

\end{document}